\documentclass[aps,prd,10pt,floatfix,nofootinbib,preprintnumbers]{revtex4-2}

\usepackage{graphicx}
\usepackage{dcolumn}
\usepackage{amsmath,amssymb,epsfig}
\usepackage{bm}
\usepackage{booktabs}
\usepackage{longtable}
\usepackage{multirow}
\usepackage{array}
\usepackage{hyperref}
\hypersetup{hidelinks}
\hypersetup{colorlinks=true,allcolors=blue}
\usepackage{url}
\graphicspath{{./}{mars_figures/}}

\allowdisplaybreaks
\newcommand{\bvec}[1]{\boldsymbol{#1}}
\newcommand{\cO}{{\cal O}}
\newcommand{\TCA}{\mathrm{TCA}}
\newcommand{\TCB}{\mathrm{TCB}}
\newcommand{\TDB}{\mathrm{TDB}}
\newcommand{\TT}{\mathrm{TT}}

\newcommand{\MCRS}{\mathrm{MCRS}}
\newcommand{\BCRS}{\mathrm{BCRS}}

\newcommand{\mars}{\mathrm{Ma}}

\newcommand{\dd}{\mathrm{d}}
\newcommand{\half}{\frac{1}{2}}

\begin{document}

\title{High-Precision Relativistic Time Scales for Mars Surface and Orbital Clocks}

\author{Slava G. Turyshev}
\affiliation{ 
Jet Propulsion Laboratory, California Institute of Technology,\\
4800 Oak Grove Drive, Pasadena, CA 91109-0899, USA
}%

\date{September 15, 2026}

\begin{abstract}
We develop a Mars-centered post-Newtonian framework for relating barycentric coordinate time, Mars-centered coordinate time, a conventional Mars surface time scale, and the proper times of landed and orbiting clocks. The construction follows the International Astronomical Union (IAU) Barycentric Celestial Reference System/Barycentric Coordinate Time (BCRS/TCB) formalism, introduces Areocentric Coordinate Time (\(\TCA\)), and writes each clock transformation as a secular rate plus zero-mean periodic terms.  Contributions are retained when their fractional-frequency amplitude exceeds \(5\times10^{-18}\) or their one-way accumulated timing amplitude exceeds \(0.1\,\mathrm{ps}\).  The numerical realization uses the GMM-3 Mars gravity field through degree and order 120, exact point-mass tidal potentials from the Sun, Phobos, and Deimos with origin and dipole terms removed, and explicit checks of stationary-model local \(c^{-4}\) terms and external-perturber scales.  Representative low-Mars-orbit, areostationary, Phobos-/Deimos-distance, and highly elliptical relay regimes are evaluated.  In leading monopole rate estimates relative to the adopted Mars surface scale, a 300 km near-polar clock is slower by \(4.56\,\mu\mathrm{s\,d^{-1}}\), while
areostationary and Deimos-distance clocks are faster by \(9.13\) and \(9.52\,\mu\mathrm{s\,d^{-1}}\), respectively.  The leading Mars-\(J_2\) timing line is about \(87\,\mathrm{ps}\) at 300 km altitude and remains several picoseconds near areostationary radius for latitude-varying areosynchronous cases; perihelion-scaled solar tides become sub-ps retained terms in high relay orbits.  A consolidated hierarchy separates mandatory, regime-dependent, and safely sub-threshold corrections.  The result is a reference-system and model-retention framework, not a final operational Mars Time Ephemeris: a realized sub-ps system still requires a selected planetary ephemeris, Mars orientation and seasonal-gravity model, spacecraft orbit-determination solution, calibrated link delays, and a covariance analysis.  In particular, time-variable low-degree gravity from seasonal CO\(_2\) exchange is a leading surface-realization term and must be modeled, monitored, or empirically bounded before sub-picosecond Mars surface-scale claims are made.
\end{abstract}

\maketitle

\section{Introduction}\label{sec:introduction}

Future Mars operations will require a timing architecture that is both relativistically well-defined and operationally usable.  Current missions can usually treat a spacecraft clock as an engineering counter calibrated to ephemeris time through a spacecraft-clock calibration and orbit-determination solution.  That practice is adequate for present radiometric navigation, but it does not by itself define a Mars-centered coordinate time, a Mars surface time scale, or a common time-transfer framework for landed clock networks, relay constellations, optical links, inter-spacecraft ranging, autonomous navigation, and future Mars Global Navigation Satellite System (GNSS)-like positioning, navigation, and timing (PNT) infrastructure.  Once clocks reach fractional stabilities near $10^{-17}$--$10^{-18}$, the timing environment at Mars is no longer represented by a single rate offset.  Solar-system barycentric dynamics, Mars' eccentric heliocentric orbit, the Martian gravity field, topography, rotation, atmospheric/seasonal mass exchange, spacecraft orbital energy, and radio/optical propagation all produce measurable contributions.

The formal foundation is the International Astronomical Union (IAU) post-Newtonian Barycentric Celestial Reference System (BCRS) and Geocentric Celestial Reference System (GCRS) framework, in which Barycentric Coordinate Time (TCB) and Geocentric Coordinate Time (TCG) are coordinate times and proper time is obtained by integrating the metric along the clock worldline \cite{Soffel2003,IAU2000,IERS2010}.  The same construction can be specialized to a Mars-centered celestial reference system.  Earlier Mars time-transfer work introduced Areocentric Coordinate Time (TCA) as a Mars local time scale analogous to TCG and treated both the proper-time--TCA and TCA--TCB links at coarser clock thresholds \cite{XuYuXie2016,YangXuYuLiuXie2017}.  Separately, recent clock-comparison work quantified the large mean and periodic proper-time difference between clocks on Mars, the Moon, and Earth, finding that Mars areoid clocks tick faster than Earth geoid clocks by hundreds of microseconds per day with a large Martian-year modulation \cite{AshbyPatla2025}.  What is still missing is an implementation-oriented Mars analogue of the high-precision lunar/cislunar timing framework: a single, stepwise derivation from reference systems to operational orbit-dependent clock budgets, light-time observables, simulations, and validation checks.

Two recent lunar and cislunar time-scale analyses provide the closest methodological precedents for the present Mars work.  Turyshev et al. derived the transformation between a lunar surface time scale (TL) and Terrestrial Time (TT), including the $56.0256\,\mu\mathrm{s}\,\mathrm{d}^{-1}$ secular lunar-surface drift, the leading $\simeq0.470\,\mu\mathrm{s}$ periodic term, and the corresponding Barycentric Dynamical Time (TDB)-compatible spatial scaling and Lorentz contraction of Moon-centered coordinates \cite{Turyshev2025Time}.  The subsequent cislunar-navigation framework extended that construction to orbit-dependent clocks, applying the explicit retention policy $5\times10^{-18}$ in fractional rate and $0.1\,\mathrm{ps}$ in one-way timing amplitude across the very-low/low lunar orbit, elliptical lunar frozen orbit, Earth--Moon $L_1$, and near-rectilinear halo orbit regimes (vLLO, LLO, ELFO, $L_1$, and NRHO) \cite{Turyshev2026Cislunar}.  Related work defines Lunar Coordinate Time (TCL) within the IAU framework, analyzes lunar-clock frequency shifts and candidate lunar reference time scales, constructs numerical lunar time ephemerides, and studies lunar reference-frame realization \cite{IAU2024,KopeikinKaplan2024,Bourgoin2025,Defraigne2025,Lu2025LTE440,Fienga2024,Sosnica2025}.  Our contribution is not a repetition of that lunar program; it transfers the same rigorous reference-system and retention-budget methodology to Mars, where the large orbital eccentricity, the GMM-3/MRO120D-class gravity field, Phobos/Deimos perturbations, seasonal mass exchange, areostationary and high-relay regimes, and Earth--Mars light-time geometry create a different accuracy hierarchy.  The relativity-specific result is the Mars-tailored retained-term hierarchy: the IAU-style BCRS--MCRS transformation through the formally relevant terms, the separation between a conventional surface coordinate scale and site proper-time realization, the exact eccentric-orbit clock term used for high relay orbits, and the static-gravity degree/order tests needed to decide which terms survive a $5\times10^{-18}$ fractional-rate or $0.1\,\mathrm{ps}$ accumulated-timing gate.

This paper supplies that construction.  We define a Mars-centered celestial reference system (MCRS) with coordinate time $T\equiv\TCA$, derive the BCRS--MCRS transformation by applying the IAU BCRS--GCRS transformation to the Mars ephemeris origin, and define a candidate Mars surface scale $T_M$ by a conventional constant-rate scaling of TCA.  The proper time of a landed or orbiting clock is then written in the form
\begin{equation}
\tau-T_{\rm ref}=\Delta_{\rm sec}(t)-P_{\rm common}(t)-P_{\rm orb}(t)-\Delta_{\rm geom}(t)+\epsilon(t),
\label{eq:auto:001}
\end{equation}
where $\Delta_{\rm sec}$ is an accumulated secular time offset, $P_{\rm common}$ contains the Earth--Mars barycentric periodic terms, $P_{\rm orb}$ is the local Mars-orbit periodic correction, $\Delta_{\rm geom}$ contains position-dependent BCRS--MCRS terms, and $\epsilon$ is the retained residual error.  The same accuracy policy is used throughout:
\begin{equation}
\epsilon_f=5\times10^{-18},\qquad \epsilon_t=0.1\,\mathrm{ps}.
\label{eq:auto:002}
\end{equation}
\begin{equation}
\bar y_{\mathcal I}=\frac{1}{T_2-T_1}\int_{T_1}^{T_2}y(T)\,\dd T,
\qquad
A_f[y;\mathcal I]=\max_{T\in\mathcal I}|y(T)-\bar y_{\mathcal I}|,
\label{eq:retention_rate_metric}
\end{equation}
\begin{equation}
P_y(T;\mathcal I)=\int_{T_1}^{T}\big[y(T')-\bar y_{\mathcal I}\big]\,\dd T',
\qquad
A_t[y;\mathcal I]=\frac{\max_{\mathcal I}P_y-\min_{\mathcal I}P_y}{2},
\label{eq:retention_timing_metric}
\end{equation}
Here $\mathcal I=[T_1,T_2]$ is the declared evaluation interval.  A secular term is retained when $|\bar y_{\mathcal I}|>\epsilon_f$; a residual term is retained when either $A_f>\epsilon_f$ or $A_t>\epsilon_t$.  Thus the timing gate is trajectory-, interval-, mean-removal-, and observable-dependent.  Root-mean-square spectra are used only for screening and do not replace the maximum-rate test in Eq.~\eqref{eq:retention_rate_metric}.
The fractional threshold corresponds to $0.432\,\mathrm{ps}$ per SI day, while $0.1\,\mathrm{ps}$ corresponds to $0.030\,\mathrm{mm}$ of one-way light-travel distance.  These thresholds are more stringent than most near-term Mars tracking systems require, but they make the truncation logic explicit and future-proof for optical clocks, optical links, and precision geodesy.

The paper is organized as follows. Section~\ref{sec:reference_systems} defines the Mars-centered reference system, the time coordinate TCA, and the BCRS--MCRS transformation.  Section~\ref{sec:time_scales} introduces
TDB, TCA, and the candidate Mars surface scale \(T_M\).  Section~\ref{sec:mars_potential} specifies the Mars gravitational, rotational, topographic, seasonal, and tidal potential model.  Section~\ref{sec:proper_time} derives the master proper-time relation for landed and orbiting clocks.  Section~\ref{sec:orbit_regimes} evaluates representative Mars orbital regimes and gives secular and periodic timing budgets.  Section~\ref{sec:numerical_simulations} describes the deterministic numerical simulations and model-retention diagnostics. Section~\ref{sec:time_transfer} treats one-way and two-way radio/optical time transfer.  Section~\ref{sec:implementation} gives an implementation workflow, and Section~\ref{sec:validation} describes validation with analytic limits, tracking data, clock data, and Earth-based time-scale comparisons.  Section~\ref{sec:discussion} discusses limitations and standardization choices, and Section~\ref{sec:conclusions} summarizes the main results.

\subsection{Scope, conventions, and accuracy philosophy}\label{subsec:scope_conventions}

\begin{table}[t]
\caption{Principal abbreviations and time-scale symbols used in this paper.}
\label{tab:abbreviations}
\begin{tabular}{ll}
\hline\hline
Abbreviation & Meaning \\
\hline
ASO & Areostationary orbit \\
BCRS & Barycentric Celestial Reference System \\
DSN & Deep Space Network \\
ELFO & Elliptical lunar frozen orbit \\
GCRS & Geocentric Celestial Reference System \\
GMM-3 & Mars gravity model GMM-3 \\
GNSS & Global Navigation Satellite System \\
HEO & Highly elliptical orbit \\
IAU & International Astronomical Union \\
JPL & Jet Propulsion Laboratory \\
LCRS & Lunicentric Celestial Reference System \\
LLO & Low lunar orbit \\
LMO & Low Mars orbit \\
MCRS & Mars-centered celestial reference system \\
MRO120D & Mars gravity model MRO120D \\
NRHO & Near-rectilinear halo orbit \\
OD & Orbit determination \\
PNT & Positioning, navigation, and timing \\
SHADR & Spherical-harmonic ASCII data record \\
SRP & Solar radiation pressure \\
TCA & Areocentric Coordinate Time; Mars-centered coordinate time used here \\
TCB & Barycentric Coordinate Time \\
TCG & Geocentric Coordinate Time \\
TCL & Lunar Coordinate Time \\
TDB & Barycentric Dynamical Time \\
TT & Terrestrial Time \\
TL & Lunar surface time scale \\
\(T_M\) & Candidate conventional Mars surface time scale \\
vLLO & Very low lunar orbit \\
WGCCRE & IAU Working Group on Cartographic Coordinates and Rotational Elements \\
\hline\hline
\end{tabular}
\end{table}

The notation in this work specifies relativistic conventions rather than a Mars civil-time standard (Table~\ref{tab:abbreviations}). TCA denotes the coordinate time of the MCRS, analogous to TCG, while $T_M$ denotes a candidate Mars surface-like coordinate scale obtained by a conventional constant-rate scaling of TCA, analogous to TT. Neither is Mars Sol Date, local mean solar time, Coordinated Mars Time, or another solar/civil-time convention. No IAU adoption of these Mars-specific names or of the defining surface-scale constant is assumed. The MCRS is kinematically non-rotating with respect to the BCRS; a Mars-fixed frame requires an adopted orientation model \cite{Archinal2018}.

The manuscript separates three levels of fidelity. The \emph{formal model} gives software-independent post-Newtonian transformations and retention criteria. The \emph{diagnostic calculations} use the constants and prescribed Keplerian or circular trajectories specified in Sections~\ref{sec:mars_potential} and~\ref{sec:numerical_simulations}. An \emph{operational realization} must instead use mutually consistent planetary and satellite ephemerides, gravity and orientation models, spacecraft dynamics, and numerical quadrature. Dependence of a numerical realization on those inputs is distinct from dependence of the theory on a software package.

\section{Reference systems and time coordinates}\label{sec:reference_systems}

\subsection{BCRS conventions and notation}\label{subsec:bcrs_conventions}

Let $(ct,\bvec{x})$ be BCRS coordinates with $t\equiv\TCB$. Greek indices $\alpha,\beta,\ldots=0,1,2,3$ label spacetime components; Latin indices $i,j,\ldots=1,2,3$ label spatial components. Repeated spatial indices are summed with $\delta_{ij}$. We adopt the IAU metric signature \cite{Soffel2003,IAU2000}
\begin{equation}
\eta_{\alpha\beta}=\mathrm{diag}(-1,+1,+1,+1),
\label{eq:auto:003}
\end{equation}
with $-c^2\dd\tau^2=g_{\alpha\beta}\dd x^\alpha\dd x^\beta$ on a timelike worldline. The BCRS metric is
\begin{align}
 g_{00}(t,\bvec{x}) &= -1+\frac{2w(t,\bvec{x})}{c^2}-\frac{2w^2(t,\bvec{x})}{c^4}+\cO(c^{-5}),\label{eq:bcrs_metric00}\\
 g_{0i}(t,\bvec{x}) &= -\frac{4w^i(t,\bvec{x})}{c^3}+\cO(c^{-5}),\\
 g_{ij}(t,\bvec{x}) &= \delta_{ij}\left(1+\frac{2w(t,\bvec{x})}{c^2}\right)+\cO(c^{-4}).
\end{align}
The scalar potential is decomposed into point-mass monopoles, extended-body multipoles, and post-Newtonian interaction terms.  The BCRS origin remains the solar-system barycenter.  The local origin must coincide with the center used by the Mars gravity field and body-fixed rotation model; in the diagnostic model below it is the Mars body center of mass.  We denote its BCRS position, velocity, and acceleration by
\begin{equation}
\bvec{x}_{\mars}(t),\qquad \bvec{v}_{\mars}=\frac{\dd\bvec{x}_{\mars}}{\dd t},\qquad \bvec{a}_{\mars}=\frac{\dd\bvec{v}_{\mars}}{\dd t},
\label{eq:auto:004}
\end{equation}
and define
\begin{equation}
\bvec{r}_{\mars}=\bvec{x}-\bvec{x}_{\mars}(t),\qquad r_{B\mars}=|\bvec{x}_{\mars}-\bvec{x}_{B}|.
\label{eq:auto:005}
\end{equation}
Here $B$ labels bodies external to that Mars body-center worldline.  An operational product must identify the ephemeris origin and planetary/satellite state combination used to construct $\bvec{x}_{\mars}$; a Mars-system barycenter must not be mixed directly with a body-centered gravity field.  The retained external-body set should include the Sun, major planetary perturbers, the Earth--Moon system, Phobos, and Deimos whenever required by the dynamics, clock, or light-time budget.

\subsection{Mars-centered celestial reference system}\label{subsec:mcrs}

We define the Mars-centered celestial reference system (MCRS) as the local, kinematically non-rotating, post-Newtonian reference system centered on the Mars ephemeris origin.  The MCRS coordinates are $(cT,\bvec{X})$ with
\begin{equation}
T\equiv\TCA.
\label{eq:auto:006}
\end{equation}
Here and throughout the paper, $T$ is TCA, \emph{not} TCG: the same letter denotes Geocentric Coordinate Time in the original IAU BCRS--GCRS transformation. The present specialization changes the reference worldline to Mars; it does not equate the two coordinate times.

Define the retained Newtonian local scalar potential by
\begin{equation}
\mathcal U_{\mars}(T,\bvec X)\equiv U_{\mars}(T,\bvec X)+U_{\rm tid}(T,\bvec X).
\end{equation}
The local metric has the IAU post-Newtonian form. In the displayed metric, $W_{\mars}=\mathcal U_{\mars}+\cO(c^{-2})$ denotes the complete scalar potential; its post-Newtonian and inertial pieces must be supplied in a full reference-system realization \cite{Soffel2003}. The local $c^{-2}$ clock budgets use $\mathcal U_{\mars}$:
\begin{align}
 G_{00}(T,\bvec{X}) &= -1+\frac{2W_{\mars}}{c^2}
 -\frac{2W_{\mars}^2}{c^4}+\cO(c^{-5}),\label{eq:mcrs_metric00}\\
 G_{0i}(T,\bvec{X}) &= -\frac{4}{c^3}W^i_{\mars}(T,\bvec{X})+\cO(c^{-5}),\\
 G_{ij}(T,\bvec{X}) &= \delta_{ij}\left(1+\frac{2W_{\mars}}{c^2}\right)+\cO(c^{-4}).
\end{align}
For the stationary monopole-plus-spin local clock estimates bounded in Sec.~\ref{subsec:master_proper_time}, the $c^{-4}$ contributions are below $5\times10^{-18}$ for the regimes treated here.  The $c^{-4}$ terms in the BCRS-to-MCRS time transformation, however, must be retained when a formal $5\times10^{-18}$ rate budget is imposed, because the solar potential and Mars barycentric velocity produce $c^{-4}$ rate terms of order $5\times10^{-17}$, or several ps/day.

The MCRS is not a body-fixed system.  A body-fixed Mars frame is obtained from the MCRS by a time-dependent rotation matrix $\mathcal{R}_{\mars}(T)$ constructed from the IAU/WGCCRE pole, prime meridian, precession, nutation, and polar-motion model.  If $\bvec{X}_{\rm bf}$ denotes a Mars-fixed vector,
\begin{equation}
\bvec{X}=\mathcal{R}_{\mars}(T)\,\bvec{X}_{\rm bf}.
\label{eq:auto:007}
\end{equation}
All gravity-field coefficients are naturally defined in the body-fixed frame, while proper-time calculations are most cleanly expressed in the MCRS.

\subsection{BCRS--MCRS coordinate transformation}\label{subsec:bcrs_mcrs}

The IAU BCRS--GCRS transformation can be specialized to any freely falling solar-system body.  To avoid ambiguity in the formally retained \(c^{-4}\) terms, we write the Mars transformation in the standard IAU form rather than as a partially expanded scalar integral.  Let \(\bar w\) and \(\bar w^i\) denote the external BCRS scalar and vector potentials, evaluated with the Mars self-field removed,
\begin{equation}
  \bar w_{\rm ext}(t,\bvec{x}_{\mars})=
  \sum_{B\ne\mars}\frac{GM_B}{r_{B\mars}}+\bar w_{\rm multipole}+\bar w_{\rm PN},
  \qquad
  \bar w^i_{\rm ext}(t,\bvec{x}_{\mars})=
  \sum_{B\ne\mars}\frac{GM_B v_B^i}{r_{B\mars}}+\cdots .
  \label{eq:external_potentials}
\end{equation}
The ellipses denote the corresponding extended-body and post-Newtonian pieces retained in the adopted ephemeris model.  With \(r_{\mars}^i=x^i-x_{\mars}^i(t)\), the time transformation is
\begin{align}
T = t &-\frac{1}{c^2}\left[A(t)+v_{\mars}^i r_{\mars}^i\right]
+\frac{1}{c^4}\left[B(t)+B^i(t)r_{\mars}^i+B^{ij}(t)r_{\mars}^i r_{\mars}^j+C(t,\bvec{x})\right]
+\cO(c^{-5}),\label{eq:tca_tcb}
\end{align}
where
\begin{align}
 \frac{\dd A}{\dd t} &= \frac{1}{2}v_{\mars}^2+\bar w_{\rm ext}(\bvec{x}_{\mars}),\nonumber\\
 \frac{\dd B}{\dd t} &=-\frac{1}{8}v_{\mars}^4
 -\frac{3}{2}v_{\mars}^2\bar w_{\rm ext}(\bvec{x}_{\mars})
 +4v_{\mars}^i\bar w^i_{\rm ext}(\bvec{x}_{\mars})
 +\frac{1}{2}\bar w_{\rm ext}^2(\bvec{x}_{\mars}),
 \label{eq:tca_AB_derivatives}
\end{align}
and
\begin{align}
B^i &=-\frac{1}{2}v_{\mars}^2v_{\mars}^i
 +4\bar w^i_{\rm ext}(\bvec{x}_{\mars})
 -3v_{\mars}^i\bar w_{\rm ext}(\bvec{x}_{\mars}),\nonumber\\
B^{ij} &=-v_{\mars}^i Q_j
 +2\partial_j\bar w^i_{\rm ext}(\bvec{x}_{\mars})
 -v_{\mars}^i\partial_j\bar w_{\rm ext}(\bvec{x}_{\mars})
 +\frac{1}{2}\delta^{ij}\frac{\partial \bar w_{\rm ext}}{\partial t}(\bvec{x}_{\mars}),\nonumber\\
C(t,\bvec{x})&=-\frac{1}{10}r_{\mars}^2\,\dot a_{\mars}^i r_{\mars}^i,
\qquad
Q_i=\partial_i\bar w_{\rm ext}(\bvec{x}_{\mars})-a_{\mars}^i .
\label{eq:tca_B_coefficients}
\end{align}
Repeated spatial indices are summed.  Equations~\eqref{eq:tca_tcb}--\eqref{eq:tca_B_coefficients} are the IAU BCRS--GCRS transformation specialized to the selected Mars origin \cite{Soffel2003,IERS2010}.  The spatial transformation used for Mars-local positions is
\begin{equation}
\bvec{X}=\bvec{r}_{\mars}+\frac{1}{c^2}\bigg\{\half(\bvec{v}_{\mars}\cdot\bvec{r}_{\mars})\bvec{v}_{\mars}+\bar w_{\rm ext}(\bvec{x}_{\mars})\bvec{r}_{\mars}
+(\bvec{a}_{\mars}\cdot\bvec{r}_{\mars})\bvec{r}_{\mars}-\half r_{\mars}^2\bvec{a}_{\mars}\bigg\}+\cO(c^{-4}).\label{eq:space_transform}
\end{equation}

At the Mars origin, the position-dependent terms vanish and Eq.~\eqref{eq:tca_tcb} reduces to \(\dd T/\dd t=1-c^{-2}\dd A/\dd t+c^{-4}\dd B/\dd t\).  Away from the origin, the large \((\bvec{v}_{\mars}\cdot\bvec{r}_{\mars})/c^2\) term and the formally retained \(c^{-4}\) spatial terms must be evaluated at the event, not absorbed into a constant rate.
For consistency with the retention accounting used elsewhere in this paper, the remainders in Eqs.~\eqref{eq:tca_tcb} and \eqref{eq:space_transform} should be interpreted as ephemeris-interval bounds, not merely formal post-Newtonian order symbols.  We write
\begin{align}
  \Delta T_{\rm rem}(t) &=
  O\!\left(c^{-5};\,\epsilon_{T}^{(5)}(t-t_0);\,\epsilon_{T}^{\rm per}\right),
  \label{eq:tca_remainder_bound}\\
  \Delta \bvec{X}_{\rm rem}(t) &=
  O\!\left(c^{-4};\,\epsilon_{X}^{(4)}\right),
  \label{eq:space_remainder_bound}
\end{align}
where \(\epsilon_T^{(5)}\) is the largest omitted fractional-rate
contribution over the adopted ephemeris interval, \(\epsilon_T^{\rm per}\) is the associated bounded periodic timing contribution, and
\(\epsilon_X^{(4)}\) is the largest omitted spatial transformation term over the spatial domain of interest.  In a Mars Time Ephemeris product these quantities should be evaluated directly from the adopted planetary and satellite ephemerides, the retained body list, and the selected multipole model, and then reported with the same \(5\times10^{-18}\) and \(0.1\,\mathrm{ps}\) retention gates
used for the local clock model.  The diagnostic simulations below quantify the corresponding local proper-time truncation terms; the BCRS--MCRS transformation residuals must be evaluated for the final ephemeris interval rather than assigned a universal Mars-independent number.

The position-dependent term $(\bvec{v}_{\mars}\cdot\bvec{r}_{\mars})/c^2$ is large.  With $v_{\mars}\simeq24.1\,\mathrm{km\,s^{-1}}$, its amplitude is $\sim0.91\,\mu\mathrm{s}$ for a surface site, $0.99\,\mu\mathrm{s}$ for a 300-km orbiter, $5.48\,\mu\mathrm{s}$ for an areostationary orbiter, and $6.30\,\mu\mathrm{s}$ at Deimos distance.  It is therefore a mandatory term in any transformation between BCRS coordinate time and local Mars coordinate time.

\subsection{Mean TCA--TCB rate}\label{subsec:tca_tcb_rate}

Define the Mars barycentric energy function at the Mars origin by
\begin{align}
\mathcal{E}_{\mars}(t)
&=\frac{1}{c^2}\left\{\frac{1}{2}v_{\mars}^2+\bar w_{\rm ext}(\bvec{x}_{\mars})\right\}
-\frac{1}{c^4}\frac{\dd B}{\dd t}\nonumber\\
&=\frac{1}{c^2}\left\{\frac{1}{2}v_{\mars}^2+\bar w_{\rm ext}(\bvec{x}_{\mars})\right\}
+\frac{1}{c^4}\left\{\frac{1}{8}v_{\mars}^4
+\frac{3}{2}v_{\mars}^2\bar w_{\rm ext}(\bvec{x}_{\mars})
-4v_{\mars}^i\bar w^i_{\rm ext}(\bvec{x}_{\mars})
-\frac{1}{2}\bar w_{\rm ext}^2(\bvec{x}_{\mars})\right\}.
\label{eq:autoalign:001}
\end{align}
We split it into a mean rate and a periodic term,
\begin{equation}
\mathcal{E}_{\mars}(t)=L_A+\dot P_A(t),\qquad \langle \dot P_A\rangle=0.\label{eq:LA_def}
\end{equation}
The final value of $L_A$ must be obtained by numerical quadrature over the same planetary ephemeris used for operations.  A useful Sun-only Keplerian estimate is
\begin{equation}
L_A^{(\odot)}\simeq\frac{3GM_\odot}{2a_{\mars}c^2}=9.72\times10^{-9}=0.8396\,\mathrm{ms}\,\mathrm{d}^{-1},\label{eq:LA_sun}
\end{equation}
where $a_{\mars}\simeq1.523679\,\mathrm{au}$.  The leading eccentricity term has fractional rate amplitude
\begin{equation}
\delta \dot P_A \simeq \frac{2GM_\odot}{a_{\mars}c^2}e_{\mars}\cos M_{\mars},
\label{eq:auto:008}
\end{equation}
which gives a simple Keplerian rate amplitude of about $1.21\times10^{-9}$, or $104.6\,\mu\mathrm{s}\,\mathrm{d}^{-1}$, and an integrated TCA--TCB annual timing amplitude of about $11\,\mathrm{ms}$.  Published numerical comparisons of Mars and Earth surface clocks include both planets' eccentric motions, the Earth--Moon system, and the adopted tidal model \cite{AshbyPatla2025}; their modulation is not the Mars-origin term alone.

For operational use it is useful to define a Mars time ephemeris by
numerically integrating the barycentric energy function, Eq.~(\ref{eq:autoalign:001}),
\begin{equation}
  I_A(t)=\int_{t_0}^{t}\mathcal{E}_{\mars}(t')\,dt',
  \label{eq:mars_time_ephemeris_integral}
\end{equation}
and decomposing it over the adopted ephemeris interval
\([t_1,t_2]\) as
\begin{align}
  L_A&=\frac{I_A(t_2)-I_A(t_1)}{t_2-t_1},\nonumber\\
  P_A(t)-P_A(t_0)&=I_A(t)-L_A(t-t_0).
  \label{eq:mars_time_ephemeris_split}
\end{align}
The additive constant in $P_A$ is conventional; setting $P_A(t_0)=0$ is convenient.  This definition gives $P_A(t_2)-P_A(t_1)=0$ over the averaging interval and recovers $\mathcal E_{\mars}=L_A+\dot P_A$ on differentiation.  The pair $\{L_A,P_A(t)\}$ is ephemeris- and interval-dependent. Mean subtraction fixes endpoint closure on that interval; it does not establish strict periodicity on an arbitrary longer interval.  The Sun-only expression in Eq.~\eqref{eq:LA_sun} is only an analytic scale check; a final $\TCA-\TCB$ product must use the same ephemeris, constants, time convention, and interpolation scheme as the orbit and light-time solution.

\section{TDB, TCA, and candidate Mars surface time scales}\label{sec:time_scales}

\subsection{TDB as the practical ephemeris argument}\label{subsec:tdb_ephemeris}

TDB is defined as a linear transformation of TCB under the IAU 2006 convention \cite{IAU2006TDB,IERS2010},
\begin{equation}
\TDB=\TCB-L_B(\TCB-T_0)+\TDB_0,
\label{eq:auto:009}
\end{equation}
with the IAU defining constants
\begin{equation}
L_B=1.550519768\times10^{-8},\qquad T_0=2443144.5003725\,\mathrm{JD},\qquad \TDB_0=-65.5\,\mu\mathrm{s}.
\label{eq:auto:010}
\end{equation}
The epoch $T_0$ is specified here by its Julian Date; time differences in Eq.~\eqref{eq:auto:009} and $\TDB_0$ are expressed in seconds. An ephemeris may use TDB-compatible quantities without changing the reference-system construction. The relevant time-scale chain is
\begin{equation}
\TCA \leftrightarrow \TCB \leftrightarrow \TDB \leftrightarrow \TT,
\label{eq:auto:011}
\end{equation}
with TCA--TCB obtained from Eq.~\eqref{eq:tca_tcb} and TDB--TCB from the IAU defining transformation.

Every state and parameter set must carry its time-scale convention. BCRS expressions use unscaled TCB-compatible quantities; Mars-local expressions use unscaled MCRS/TCA quantities unless a scaling is stated explicitly. A state supplied in a TDB-compatible representation is not inserted unchanged into a TCB expression. Following Klioner \cite{Klioner2008}, a constant scaling of coordinate time is accompanied by consistent scalings of spatial coordinates and gravitational mass parameters so that the dynamical equations retain their form. With $F_B\equiv1-L_B$,
\begin{align}
  \TDB&=\TCB-L_B(\TCB-T_0)+\TDB_0,\nonumber\\
  \bvec{x}_{\TDB}&=F_B\bvec{x}_{\TCB},\qquad
  (GM)_{\TDB}=F_B(GM)_{\TCB},\nonumber\\
  \bvec v_{\TDB}&=\bvec v_{\TCB},\qquad
  \bvec a_{\TDB}=F_B^{-1}\bvec a_{\TCB}.
  \label{eq:tdb_compatible_scaling}
\end{align}
Velocity is invariant under this constant scaling because position and time carry the same factor; this statement is not a velocity transformation between the BCRS and MCRS. The scaled quantities are distinct coordinate quantities expressed in the same SI units, not values in a separate system of TDB-specific units \cite{Klioner2008}.  A practical implementation should use one compatible set of states, mass parameters, and independent variables throughout; the distinction is not negligible when secular rates are specified at the $10^{-18}$ level.

\subsection{Definition of a Mars surface time scale}\label{subsec:mars_surface_scale}

A Mars surface time scale should be distinguished from Mars solar time, Mars Sol Date, and Coordinated Mars Time in the civil/solar sense \cite{Mars24}.  Solar time is tied to rotation relative to the Sun; the relativistic scale $T_M$ used here is a coordinate-time convention.

Let $T\equiv\TCA$.  Define the provisional Mars surface scale by the exact constant-rate relation
\begin{equation}
\frac{\dd T_M}{\dd\TCA}=1-L_{\rm surf}^{\rm def},
\label{eq:TM_def_rate}
\end{equation}
where $L_{\rm surf}^{\rm def}$ is conventional.  Its associated defining potential is
\begin{equation}
W_{\rm def}\equiv c^2L_{\rm surf}^{\rm def}.
\label{eq:Lsurf_def}
\end{equation}
For the diagnostic calculations we adopt
\begin{equation}
L_{\rm surf}^{\rm def}=1.406355\times10^{-10}
=12.15091\,\mu\mathrm{s\,d^{-1}},
\label{eq:Lsurf_def_value}
\end{equation}
where the displayed digits define the scale rather than claim a geophysical determination at that precision.

Separately, let $W_0^{\mars}$ be the potential assigned to a selected reference areoid under a stated static gravity and rotation convention, and define
\begin{equation}
L_0^{\mars}\equiv\frac{W_0^{\mars}}{c^2},\qquad
\delta L_0\equiv L_0^{\mars}-L_{\rm surf}^{\rm def}
=\frac{W_0^{\mars}-W_{\rm def}}{c^2}.
\label{eq:Lsurf_convention}
\end{equation}
The offset $\delta L_0$ belongs to the physical realization of $T_M$, not to its definition.  A simple equatorial degree-2 estimate is
\begin{equation}
  L_0^{(0+J_2+\Omega)}\simeq
  \frac{1}{c^2}\left\{
  \frac{GM_{\mars}}{R_0}
  +\frac{1}{2}\frac{GM_{\mars}J_{2\mars}}{R_0}
  +\frac{1}{2}\Omega_{\mars}^2R_0^2
  \right\},
  \label{eq:Lsurf_degree2_check}
\end{equation}
which gives $L_0^{(0+J_2+\Omega)}\simeq1.4078057\times10^{-10}$ at the diagnostic radius $R_0$; it is not a geophysical determination of the areoid. The difference from Eq.~\eqref{eq:Lsurf_def_value} is about $1.45\times10^{-13}$, or $12.5\,\mathrm{ns\,d^{-1}}$, and therefore must be carried explicitly through $\delta L_0$.

The analogous global scaling, with the same quantity-versus-unit distinction \cite{Klioner2008}, is
\begin{equation}
  T_M=T_{M0}+(1-L_{\rm surf}^{\rm def})(\TCA-T_{\mathrm{CA},0}),\qquad
  \bvec{X}_{M}=(1-L_{\rm surf}^{\rm def})\bvec{X}_{\TCA},\qquad
  (GM)_{M}=(1-L_{\rm surf}^{\rm def})(GM)_{\TCA}.
\label{eq:mars_surface_scaling_def}
\end{equation}
Here $T_{\mathrm{CA},0}$ and $T_{M0}$ are the assigned TCA and $T_M$ readings at one reference event. Neither $\delta L_0$ nor site, tide, seasonal, or gravity-model corrections enter this global definition; they enter only when a physical clock is related to $T_M$.

\subsection{Surface site offsets}\label{subsec:surface_site_offsets}

For a landed clock define the retained effective potential
\begin{equation}
W_{\rm site}(T)\equiv U_{\mars}(T,\bvec X_{\rm site})+U_{\rm tid}(T,\bvec X_{\rm site})+\frac{1}{2}V_{\rm site}^2(T).
\label{eq:site_effective_potential}
\end{equation}
Its convention-relative offset is
\begin{align}
\Delta W_{\rm site}^{\rm def}(T)&\equiv W_{\rm site}(T)-W_{\rm def}\nonumber\\
&=c^2\delta L_0+\big[W_{\rm site}(T)-W_0^{\mars}\big].
\label{eq:site_offset_def}
\end{align}
Thus
\begin{equation}
\frac{\dd\tau_{\rm site}}{\dd T_M}-1
\simeq-\frac{\Delta W_{\rm site}^{\rm def}(T)}{c^2}
=-\delta L_0-\frac{W_{\rm site}(T)-W_0^{\mars}}{c^2}+\cO(c^{-4}).
\label{eq:auto:014}
\end{equation}
A clock on the selected areoid is rate-matched to $T_M$ only if $W_0^{\mars}=W_{\rm def}$.

For physical height $h>0$ outward from the adopted areoid,
\begin{equation}
W_{\rm site}-W_0^{\mars}\simeq-g_{\rm eff}h+\Delta W_{\rm harm}+\Delta W_{\rm tide}+\Delta W_{\rm seasonal}+\Delta W_{\rm local},
\label{eq:site_height_potential}
\end{equation}
so the height contribution makes the elevated clock run faster than an areoid clock.  One meter corresponds to
\begin{equation}
\frac{g_{\mars}(1\,\mathrm{m})}{c^2}\simeq4.13\times10^{-17}=3.57\,\mathrm{ps}\,\mathrm{d}^{-1}.
\label{eq:auto:015}
\end{equation}
The $5\times10^{-18}$ rate target therefore corresponds to $0.12\,\mathrm{m}$ of gravitational height, and $0.1\,\mathrm{ps}$ accumulated over one day to about $2.8\,\mathrm{cm}$.  Topography must remain a site-realization correction, not part of the global definition of $T_M$.

\subsection{Mean relation to TT}\label{subsec:mean_relation_tt}

At the Mars origin, the rate relative to terrestrial TT is obtained by composing the transformations before averaging:
\begin{align}
\left\langle\frac{\dd T_M}{\dd\TT}\right\rangle-1
&=\frac{1-L_{\rm surf}^{\rm def}}{1-L_B}
\left\langle[1-\mathcal E_{\mars}(t)]\frac{\dd\TDB}{\dd\TT}\right\rangle-1
\nonumber\\
&\simeq L_B-L_A-L_{\rm surf}^{\rm def}.
\label{eq:TM_TT_mean}
\end{align}
The first line is exact in the constant scale factors at the retained order of the TCA--TCB transformation; $\dd\TDB/\dd\TT$ is supplied by the adopted Earth time ephemeris. The second line is only a leading-order scale estimate, neglecting products of small rate terms and interval-dependent residual averages. Those products must not be dropped in a $5\times10^{-18}$ implementation. Away from the Mars origin the event-dependent terms of Eq.~\eqref{eq:tca_tcb} also enter. The Sun-only estimate gives
\begin{equation}
L_B-L_A^{(\odot)}-L_{\rm surf}^{\rm def}
\simeq5.65\times10^{-9}
\simeq487.9\,\mu\mathrm{s}\,\mathrm{d}^{-1},
\label{eq:sun_only_mars_tt_check}
\end{equation}
which is only a scale check.

A physical clock on the selected reference areoid, with proper time $\tau_0$, is related to this coordinate scale by
\begin{equation}
\Big\langle\frac{\dd\tau_0}{\dd\TT}\Big\rangle-1
\simeq
\Big\langle\frac{\dd T_M}{\dd\TT}\Big\rangle-1-\delta L_0.
\label{eq:areoid_tt_bridge}
\end{equation}
The numerical comparison in Ref.~\cite{AshbyPatla2025} reports a physical Mars-areoid advance of $477.60\,\mu\mathrm{s\,d^{-1}}$, an annual rate amplitude of $226.8\,\mu\mathrm{s\,d^{-1}}$, and an additional $\sim40\,\mu\mathrm{s\,d^{-1}}$ modulation over approximately seven synodic cycles.  These values are an external validation datum, not a direct numerical realization of $T_M-\TT$, unless the reference potential and averaging convention are mapped through Eq.~\eqref{eq:areoid_tt_bridge}.  An operational Mars Time Ephemeris should integrate Eq.~\eqref{eq:tca_tcb} with one consistent planetary ephemeris and report $L_A$, $P_A(t)$, $T_M-\TT$, and the convention-to-areoid bridge separately.

\section{Mars gravitational potential model}\label{sec:mars_potential}

\subsection{Static field}\label{subsec:static_field}

Outside Mars, the Newtonian self-potential in the body-fixed frame is written
\begin{align}
U_{\mars}(r,\theta,\lambda) &= \frac{GM_{\mars}}{r}\bigg\{1+\sum_{\ell=2}^{\ell_{\max}}\sum_{m=0}^{\ell}\Big(\frac{R_0}{r}\Big)^{\ell}\bar P_{\ell m}(\cos\theta)
\Big[\bar C_{\ell m}\cos m\lambda+\bar S_{\ell m}\sin m\lambda\Big]\bigg\},\label{eq:mars_gravity}
\end{align}
Here $\theta$ is body-fixed colatitude, $\phi=\pi/2-\theta$ is latitude, and $\lambda$ follows the longitude convention of the coefficient file.  The associated Legendre functions and coefficients must use the same fully normalized geodetic convention; for the GMM-3 values in Table~\ref{tab:constants},
\begin{equation}
J_{2\mars}=-\sqrt{5}\,\bar C_{20},
\end{equation}
which provides a direct normalization check.  GMM-3 and MRO120D extend to degree and order 120 \cite{Genova2016,Konopliv2016,PGDAGMM3}; the retained degree for time transfer must be selected from the maximum-rate and finite-arc timing residuals, not from formal degree alone:
\begin{equation}
\frac{|\Delta U|}{c^2}\le5\times10^{-18},\qquad |\Delta U|\le0.45\,\mathrm{m^2\,s^{-2}}.
\label{eq:auto:016}
\end{equation}
For low orbit the degree/order-120 field is used as the diagnostic reference reconstruction even when analytic estimates retain only $J_2$, $C_{22}$, and $S_{22}$.  Near-surface clocks additionally require topography and a stated areoid convention.  Table~\ref{tab:constants} collects the analytic constants; operational software must use one internally consistent ephemeris, gravity, and rotation data set. In particular, the GMM-3 product label specifies a model-specific reference frame, explicitly not the IAU Mars frame \cite{GMM3PDSLabel}; an operational combination must map the coefficient frame and the adopted orientation consistently.

\begin{table*}[!tbp]
\caption{Representative Mars constants used for the quantitative estimates.  Values are rounded for analytic budgeting; operational software should use a single internally consistent ephemeris, gravity field, and orientation model.}
\begin{ruledtabular}
\begin{tabular}{lll}
Quantity & Symbol & Adopted value / comment\\
\hline
Speed of light & $c$ & $299792458\,\mathrm{m\,s^{-1}}$\\
TDB scaling & $L_B$ & $1.550519768\times10^{-8}$\\
Mars gravitational parameter & $GM_{\mars}$ & $4.2828372854187757\times10^{13}\,\mathrm{m^3\,s^{-2}}$\\
Mars equatorial radius & $R_{\rm eq}$ & $3396.19\,\mathrm{km}$\\
Mars gravity-field reference radius & $R_0$ & $3396.0\,\mathrm{km}$ for GMM-3/MRO120D class fields\\
Mars mean radius & $R_{\rm mean}$ & $3389.50\,\mathrm{km}$\\
Mars sidereal rotation period & $P_{\rm rot}$ & $1.02595676\,\mathrm{d}$\\
Mars rotation rate & $\Omega_{\mars}$ & $7.088218\times10^{-5}\,\mathrm{s^{-1}}$\\
Mars semimajor axis & $a_{\mars}$ & $1.523679\,\mathrm{au}$\\
Mars eccentricity & $e_{\mars}$ & $0.0934$\\
Mars degree-2 zonal & $J_{2\mars}$ & $1.9566\times10^{-3}$\\
GMM-3 normalized $\bar C_{20}$ & $\bar C_{20}$ & $-8.7502113\times10^{-4}$\\
GMM-3 normalized $\bar C_{22},\bar S_{22}$ & $\bar C_{22},\bar S_{22}$ & $-8.4635904\times10^{-5}$, $4.8934626\times10^{-5}$\\
Mars surface scaling, defining value & \(L_{\rm surf}^{\rm def}\) &
\(1.406355\times10^{-10}=12.15091\,\mu\mathrm{s\,d^{-1}}\) (provisional convention)\\
Areostationary radius & $r_{\rm ASO}$ & $20427.7\,\mathrm{km}$, altitude $17031.5\,\mathrm{km}$\\
Phobos orbital radius & $a_P$ & $9376\,\mathrm{km}$ from Mars center\\
Deimos orbital radius & $a_D$ & $23463\,\mathrm{km}$ from Mars center\\
Mars Love number & $k_2$ & $\simeq0.1697$ from GMM-3 class analyses\\
\end{tabular}
\end{ruledtabular}
\label{tab:constants}
\end{table*}

\subsection{Rotation, topography, and the areoid}\label{subsec:rotation_topography_areoid}

For a selected static gravity field and rotation convention define
\begin{equation}
W_{\rm stat}(\bvec X)\equiv U_{\mars}^{(0)}(\bvec X)+\frac{1}{2}|\bvec\Omega_{\mars}\times\bvec X|^2,
\label{eq:auto:017}
\end{equation}
and let the reference areoid satisfy $W_{\rm stat}(\bvec X_{\rm areoid})=W_0^{\mars}$.  Time-variable tides, seasonal mass exchange, atmospheric loading, and local signals are realization terms in $W_{\rm site}(T)$, not part of a permanently fixed level surface.

The areoid is not the physical topographic surface.  Mars relief of order $20\,\mathrm{km}$ can produce
\begin{equation}
\frac{g_{\mars}(20\,\mathrm{km})}{c^2}\sim8\times10^{-13}\simeq70\,\mathrm{ns}\,\mathrm{d}^{-1},
\label{eq:auto:018}
\end{equation}
so individual landed clocks require explicit geodetic and time-variable potential corrections to realize $T_M$.

\subsection{Seasonal gravity and tides}\label{subsec:seasonal_tides}

Mars has time-variable gravity from seasonal CO\(_2\) exchange, atmospheric loading, and solid-body response \cite{Genova2016}.  The static-plus-time-variable model should be written
\begin{equation}
  \bar C_{\ell m}(T)=\bar C_{\ell m}^{(0)}
  +\Delta\bar C_{\ell m}^{\rm seas}(T)
  +\Delta\bar C_{\ell m}^{\rm tide}(T),
  \qquad
  \bar S_{\ell m}(T)=\bar S_{\ell m}^{(0)}
  +\Delta\bar S_{\ell m}^{\rm seas}(T)
  +\Delta\bar S_{\ell m}^{\rm tide}(T).
  \label{eq:seasonal_gravity_model}
\end{equation}
The low-degree seasonal terms, especially \(\Delta C_{20}\),
\(\Delta C_{30}\), and related zonals, are small in instantaneous fractional rate but can accumulate to the \(0.1\,\mathrm{ps}\) level over days to weeks. The deterministic simulations in Sec.~\ref{sec:numerical_simulations} use the static GMM-3 field.  They therefore test the static-gravity truncation and the relativistic clock model, but they do not bound the error from seasonal
CO\(_2\) exchange, atmospheric loading, or other time-variable low-degree harmonics.  In an operational Mars Time Ephemeris these terms should be modeled explicitly or propagated into the clock error budget as \(\delta q_{\rm seas}=\Delta U_{\rm seas}/c^2\) and
\(\Delta\tau_{\rm seas}=-\int\delta q_{\rm seas}\,\dd T\).  A realized
sub-picosecond Mars surface time scale should not be claimed unless the seasonal contribution is either modeled or bounded below the adopted retention threshold.  The principal realization-limited contributions not bounded by the static deterministic simulations are summarized in Table~\ref{tab:realization_error_budget}.

\begin{table*}[t]
\centering
\caption{Representative realization and covariance-limited error scales not bounded by the static, deterministic GMM-3 simulations.  The entries are order-of-magnitude conversions for error-budget construction, not a mission covariance.}
\label{tab:realization_error_budget}
\renewcommand{\arraystretch}{1.05}
\begin{tabular}{ll}
\hline\hline
\parbox[t]{0.26\textwidth}{\raggedright Source or unmodeled quantity} &
\parbox[t]{0.68\textwidth}{\raggedright Representative scale and induced clock effect}\\
\hline
\parbox[t]{0.26\textwidth}{\raggedright Site height or areoid error} &
\parbox[t]{0.68\textwidth}{\raggedright A radial error of $0.10\,\mathrm{m}$ gives $4.1\times10^{-18}$, or $0.36\,\mathrm{ps}$ accumulated in one SI day; $0.028\,\mathrm{m}$ gives $0.1\,\mathrm{ps}$ in one day.}\\[0.8ex]
\parbox[t]{0.26\textwidth}{\raggedright Static gravity residual} &
\parbox[t]{0.68\textwidth}{\raggedright A residual potential $|\Delta U|=0.45\,\mathrm{m^2\,s^{-2}}$ approximates the $5\times10^{-18}$ fractional-rate gate ($c^2\epsilon_f=0.44937759\,\mathrm{m^2\,s^{-2}}$) and accumulates $0.432\,\mathrm{ps}$ in one day.}\\[0.8ex]
\parbox[t]{0.26\textwidth}{\raggedright Seasonal low-degree gravity} &
\parbox[t]{0.68\textwidth}{\raggedright A coefficient-scale error $|\Delta\bar C_{20}|\sim10^{-8}$ corresponds to a surface fractional-rate scale of order $10^{-18}$--$10^{-17}$, depending on latitude and normalization; it must be modeled or carried as a systematic.}\\[0.8ex]
\parbox[t]{0.26\textwidth}{\raggedright 300 km LMO radial OD error} &
\parbox[t]{0.68\textwidth}{\raggedright For the circular-orbit monopole clock factor, $1\,\mathrm{m}$ radial error gives $3GM_{\mars}/(2r^2c^2)\simeq5.2\times10^{-17}$, or $4.5\,\mathrm{ps}$ per day.}\\[0.8ex]
\parbox[t]{0.26\textwidth}{\raggedright Mars orientation or station-coordinate error} &
\parbox[t]{0.68\textwidth}{\raggedright The radial component maps directly to the height line above; horizontal components enter through the harmonic and centrifugal potential gradients and must be propagated with the adopted frame covariance.}\\
\hline\hline
\end{tabular}
\end{table*}

The external tidal potential admits the local expansion
\begin{equation}
U_{\rm tid}(T,\bvec{X})=\sum_{B\ne\mars}\sum_{\ell=2}^{N_B}\frac{GM_B}{r_{B\mars}}\Big(\frac{X}{r_{B\mars}}\Big)^{\ell}P_{\ell}(\hat{\bvec{n}}_{B\mars}\cdot\hat{\bvec{X}}).\label{eq:tide}
\end{equation}
For each source, the infinite series underlying this expression converges only for $X<r_{B\mars}$. Near and beyond a moon's orbital radius, Eq.~\eqref{eq:exact_tide_model}, rather than an exterior-domain use of this series, is required.

At mean heliocentric distance the Mars-surface solar quadrupole scale is only $\sim1.4\times10^{-18}$ in fractional rate.  Because the solar quadrupole tide scales as $R_{\mars\odot}^{-3}$, Mars' orbital eccentricity gives the perihelion multiplier
\begin{equation}
f_{\rm peri}\equiv\left(\frac{a_{\mars}}{r_{\rm peri}}\right)^3=(1-e_{\mars})^{-3}\simeq1.342 .
\label{eq:solar_tide_perihelion_factor}
\end{equation}
Thus the same surface scale reaches $\sim1.9\times10^{-18}$ at perihelion.  The mean-distance values grow as $X^2$, becoming $5.2\times10^{-17}$ at areostationary radius and $6.9\times10^{-17}$ at Deimos distance; the corresponding perihelion values are $7.0\times10^{-17}$ and $9.3\times10^{-17}$.  Mean-distance values are useful diagnostic scales, but the perihelion-scaled values set the retain/discard decision whenever a term can exceed the gate over the operational interval.  Solar tides are therefore negligible for circular LMO timing amplitudes but mandatory for high Mars relay orbits at the $0.1\,\mathrm{ps}$ level.  The direct Phobos and Deimos tide on Mars surface clocks is small, but the self-potentials of Phobos and Deimos are important for close moon-proximity operations: $GM_P/(100\,\mathrm{km}\,c^2)\simeq7.9\times10^{-17}$ and $GM_D/(100\,\mathrm{km}\,c^2)\simeq1.1\times10^{-17}$.

\section{Proper time of Mars surface and orbital clocks}\label{sec:proper_time}

\subsection{Master proper-time relation in the MCRS}\label{subsec:master_proper_time}

For a clock with MCRS coordinate velocity $\bvec{V}=\dd\bvec{X}/\dd T$, the convention $-c^2\dd\tau^2=G_{\alpha\beta}\dd X^\alpha\dd X^\beta$ gives
\begin{equation}
\frac{\dd\tau}{\dd T}=1-\frac{1}{c^2}\Big\{\half V^2+U_{\mars}(T,\bvec{X})+U_{\rm tid}(T,\bvec{X})\Big\}+\cO(c^{-4}).\label{eq:proper_tca}
\end{equation}
At a 300-km circular Mars orbit, $V\simeq3.40\,\mathrm{km\,s^{-1}}$ and $GM_{\mars}/r\simeq1.16\times10^7\,\mathrm{m^2\,s^{-2}}$. For the stationary monopole-plus-spin approximation, the algebraic $c^{-4}$ clock terms obey
\begin{equation}
 |\delta^{\rm alg}_{c^{-4}}|\le\frac{1}{c^4}
 \left\{\frac18 V^4+\frac32 V^2|\mathcal U_{\mars}|+
 \frac12|\mathcal U_{\mars}|^2+4|W^i_{\mars}V_i|\right\}.
 \label{eq:c4_local_bound_main}
\end{equation}
The full remainder additionally contains $-(W_{\mars}-\mathcal U_{\mars})/c^2$ and any omitted time-dependent or inertial pieces of the formal local potential. Equation~\eqref{eq:c4_local_bound_main} is therefore a bound on the specified stationary-model contributions, not a substitute for completing that potential in an operational reference system. With the conservative spin estimate $4|\bvec W_{\mars}\!\cdot\!\bvec V|\le2GM_{\mars}R_0^2\Omega_{\mars}V/r^2$, corresponding to $|\bvec S|\le M_{\mars}R_0^2\Omega_{\mars}$ in the diagnostic rotating-body model, their fractional-rate bounds are below $4\times10^{-20}$ for the 300 km LMO and $6\times10^{-20}$ for the HEO. A rate bound $\beta$ implies $|\Delta_{c^{-4}}|\le\beta|T-T_0|$ for those terms: less than $0.00082\,\mathrm{ps}$ over three LMO revolutions and $0.0025\,\mathrm{ps}$ over one HEO revolution. Longer arcs require a new accumulated-time bound.

We split the bracket in Eq.~\eqref{eq:proper_tca} into a mean and a periodic part,
\begin{equation}
\frac{1}{c^2}\Big\{\half V^2+U_{\mars}+U_{\rm tid}\Big\}=L_{\rm orb}+\dot P_{\rm orb}(T),\qquad \langle\dot P_{\rm orb}\rangle=0.\label{eq:Lorb_def}
\end{equation}
For an orbiter this integrates to
\begin{align}
\tau(T)-T_M(T)={}&\tau(T_0)-T_M(T_0)
+\left(L_{\rm surf}^{\rm def}-L_{\rm orb}\right)(T-T_0)\nonumber\\
&-\big[P_{\rm orb}(T)-P_{\rm orb}(T_0)\big]+\Delta_{c^{-4}}(T;T_0),
\label{eq:auto:020}
\end{align}
where $\Delta_{c^{-4}}$ is the integrated local remainder; its specified stationary-model part is bounded as described after Eq.~\eqref{eq:c4_local_bound_main}.  Site corrections for landed clocks are given by Eq.~\eqref{eq:auto:014}; BCRS--MCRS event terms and propagation delays enter only when timestamps or observables are transformed, not in the intrinsic proper-time integral of one clock.  Accordingly, Eq.~\eqref{eq:auto:001} is an observable-level bookkeeping identity.

Relative to TT, the constant scale factors should again be composed rather than expanded prematurely:
\begin{equation}
\frac{\dd\tau}{\dd\TT}-1=
\frac{1-L_{\rm orb}-\dot P_{\rm orb}(T)+\delta_{c^{-4}}(T)}{1-L_{\rm surf}^{\rm def}}
\frac{\dd T_M}{\dd\TT}-1.
\label{eq:tau_tt_master}
\end{equation}
Here $\delta_{c^{-4}}$ denotes the complete local remainder at the chosen model order. A numerical TT rate requires an ephemeris-specific $T_M-\TT$ product; the orbit table below therefore gives only leading local rate estimates relative to $T_M$.

\subsection{Circular orbit secular rate}\label{subsec:circular_rate}

For a circular Mars orbit of radius $r$, neglecting harmonics and tides,
\begin{equation}
V^2=\frac{GM_{\mars}}{r},\qquad L_{\circ}=\frac{3GM_{\mars}}{2rc^2}.\label{eq:L_circ}
\end{equation}
The secular offset relative to the conventionally defined surface scale $T_M$, per unit TCA, is
\begin{equation}
\Delta\dot\tau_{\circ/{\rm surf}}=L_{\rm surf}^{\rm def}-L_{\circ}.
\label{eq:auto:021}
\end{equation}
The sign is important: a low Mars orbiter can be slower than $T_M$ because of its high velocity, whereas high relay orbits are faster because both kinetic and potential terms are smaller.

The zero of Eq.~\eqref{eq:auto:021} defines a \emph{time-aligned orbit}, in the sense of the construction of Yang et al.~\cite{Yang2026TimeAligned}. In the circular monopole model,
\begin{equation}
 r_{\rm TA}=\frac{3GM_{\mars}}{2c^2L_{\rm surf}^{\rm def}}
           =\frac{3GM_{\mars}}{2W_{\rm def}},\qquad
 h_{\rm TA}=r_{\rm TA}-R_{\rm eq}.
 \label{eq:time_aligned_radius}
\end{equation}
The adopted constants give $r_{\rm TA}=5082.6\,\mathrm{km}$, $h_{\rm TA}=1686.4\,\mathrm{km}$, $r_{\rm TA}/R_{\rm eq}=1.49656$, and period $3.0559\,\mathrm h$. For a non-rotating spherical reference body with $W_{\rm def}=GM/R$, the radius is exactly $3R/2$: $1.5R$ is a center-to-orbit distance, not an altitude. The crossing is marked in Fig.~\ref{fig:circular_budget_full}. The mean rate increases monotonically through it, with local sensitivity $\partial\Delta\dot\tau/\partial r=L_{\rm surf}^{\rm def}/r_{\rm TA}\simeq2.77\times10^{-17}\,\mathrm{m^{-1}}$; a $5\times10^{-18}$ tolerance corresponds to approximately $0.18$ m in this idealized one-parameter model, not to a complete orbit-determination requirement.

For the retained local model the corresponding condition is instead
\begin{equation}
 \left\langle\frac12V^2+U_{\mars}+U_{\rm tid}\right\rangle_{\mathcal I}=W_{\rm def},
 \label{eq:time_aligned_full}
\end{equation}
with the trajectory and averaging interval $\mathcal I$ specified. Harmonics affect both the potential and the orbit; adding their direct potential to a prescribed Keplerian trajectory does not include the orbit-mediated correction. Alignment with a physical reference-areoid clock replaces $W_{\rm def}$ by $W_0^{\mars}$, together with the retained time-dependent site terms. Yang et al. report an equatorial \emph{mean semimajor axis} of $5087.696$ km for Mars under their reference-potential and $J_2$ conventions \cite{Yang2026TimeAligned}; this is not the same quantity or approximation as the monopole radius in Eq.~\eqref{eq:time_aligned_radius}. Time alignment cancels a secular drift, not the periodic correction $P_{\rm orb}$, an initial clock offset, or the convention-to-areoid difference $\delta L_0$.

\subsection{Elliptical orbit secular and periodic terms}\label{subsec:elliptic_terms}

For a prescribed Keplerian ellipse with semimajor axis $a$ and eccentricity $e$,
\begin{equation}
\Big\langle\half V^2\Big\rangle=\frac{GM_{\mars}}{2a},\qquad \Big\langle\frac{GM_{\mars}}{r}\Big\rangle=\frac{GM_{\mars}}{a},
\label{eq:auto:022}
\end{equation}
so that
\begin{equation}
L_e=\frac{3GM_{\mars}}{2ac^2}+\Big\langle U_{\rm harm}+U_{\rm tid}\Big\rangle/c^2.
\label{eq:auto:023}
\end{equation}
The zero-mean kinematic-plus-monopole part is
\begin{equation}
\dot P_{K+M}(T)=\frac{2GM_{\mars}}{c^2}\left(\frac{1}{r(T)}-\frac{1}{a}\right),
\label{eq:elliptic_rate}
\end{equation}
and, between arbitrary epochs, integrates to
\begin{equation}
P_{K+M}(T)-P_{K+M}(T_0)=\frac{2GM_{\mars}}{a c^2 n}\,e\big[\sin E(T)-\sin E(T_0)\big],
\label{eq:elliptic_exact}
\end{equation}
where $n=\sqrt{GM_{\mars}/a^3}$.  Equivalently,
\begin{align}
P_{K+M}(T)-P_{K+M}(T_0)={}&\frac{2GM_{\mars}}{a c^2 n}
\sum_{k=1}^{\infty}\frac{2}{k}\mathsf J_k(ke)\nonumber\\
&\times\big[\sin kM(T)-\sin kM(T_0)\big].
\label{eq:elliptic_series}
\end{align}
For an epoch at periapsis, the cubic expansion is
\begin{equation}
\frac{2GM_{\mars}}{a c^2 n}\left[
\left(e-\frac{e^3}{8}\right)\sin M+\frac{e^2}{2}\sin2M+\frac{3e^3}{8}\sin3M+\cO(e^4)\right].
\label{eq:elliptic_small_e}
\end{equation}
Equations~\eqref{eq:auto:023} and~\eqref{eq:elliptic_exact} use a prescribed Keplerian state; an operational perturbed trajectory must be repropagated and its kinetic term evaluated consistently. High-eccentricity relay orbits require Eq.~\eqref{eq:elliptic_exact} or direct quadrature; Eq.~\eqref{eq:elliptic_small_e} is only a low-eccentricity check.

\section{Representative Mars orbital regimes}\label{sec:orbit_regimes}

\subsection{Mars surface and landed clocks}\label{subsec:surface_landed}

A landed clock realizes proper time, not $T_M$.  For a real lander,
\begin{align}
\frac{\dd\tau_{\rm lander}}{\dd T_M}-1
&=-\frac{W_{\rm lander}(T)-W_{\rm def}}{c^2}+\cO(c^{-4})\nonumber\\
&=-\delta L_0-\frac{W_{\rm lander}(T)-W_0^{\mars}}{c^2}+\cO(c^{-4}).
\label{eq:auto:024}
\end{align}
An areoid clock is therefore rate-matched to $T_M$ only when $\delta L_0=0$.  A lander at positive physical height runs faster than an areoid clock to leading order, but its rate relative to $T_M$ also contains the convention offset $-\delta L_0$.  The dominant realization terms are topographic and local potential, rotation, time-variable low-degree gravity, loading, and tides.

For surface-network operations the offset of site $i$ is
\begin{equation}
\Delta\tau_i^{(M)}(T)=-\frac{1}{c^2}\int_{T_0}^{T}\big[W_i(T')-W_{\rm def}\big]\,\dd T'+\Delta\tau_i^{(M)}(T_0).
\label{eq:auto:025}
\end{equation}
The integrand must include the convention-to-areoid offset, topography, local gravity anomalies, Mars orientation, seasonal gravity, solid tides, atmospheric loading, and the site's actual coordinate velocity.  The common $W_{\rm def}$ cancels in a direct comparison of two sites, but not when either clock is referred to $T_M$.

\subsection{Low Mars orbit}\label{subsec:lmo}

For a circular, near-polar low Mars orbit (LMO) at altitude $h=300$--$400\,\mathrm{km}$,
\begin{equation}
 r_{\rm LMO}=R_{\rm eq}+h,
\label{eq:auto:026}
\end{equation}
with period $1.90$--$1.97\,\mathrm{h}$ and speed $3.36$--$3.40\,\mathrm{km\,s^{-1}}$.  The mean local clock factor is
\begin{equation}
L_{\rm LMO}=\frac{3}{2}\frac{GM_{\mars}}{r_{\rm LMO}c^2}+\Big\langle U_{J_2}+U_{22}+\cdots\Big\rangle/c^2.
\label{eq:auto:027}
\end{equation}
At 300 km, the monopole part is $L_{\circ}\simeq1.93\times10^{-10}=16.71\,\mu\mathrm{s}\,\mathrm{d}^{-1}$, giving the leading offset $-4.56\,\mu\mathrm{s}\,\mathrm{d}^{-1}$ relative to $T_M$. The harmonic mean in Eq.~\eqref{eq:auto:027} must still be included for a precision rate.  Its rate relative to TT must be formed from Eq.~\eqref{eq:tau_tt_master} using an ephemeris-specific $T_M-\TT$ transformation.

The $J_2$ periodic line for a circular polar orbit has the form
\begin{equation}
P_{J_2}(T)\simeq -\frac{3}{8}\frac{GM_{\mars}R_0^2J_{2\mars}}{c^2r^3n}\sin\Big(2nT+\varphi_{J2}\Big),\label{eq:J2_line}
\end{equation}
where $n=\sqrt{GM_{\mars}/r^3}$.  Numerically, $|P_{J_2}|\simeq87\,\mathrm{ps}$ at 300 km and $83\,\mathrm{ps}$ at 400 km, corresponding to one-way light ranges of $26$ and $25\,\mathrm{mm}$.  The $C_{22}$ and $S_{22}$ terms produce additional degree-2 lines and sidebands of order tens of ps, depending on inclination, longitude, and normalization convention.  Higher-degree Mars harmonics can exceed $0.1\,\mathrm{ps}$ and should be evaluated with the full degree-120 field in the operational model.  Solar tides in circular LMO are below $10^{-3}\,\mathrm{ps}$ in integrated timing amplitude and can be neglected at this threshold unless long-arc averaging of rate residuals is being studied.

\subsection{Medium relay orbit near 1000 km altitude}\label{subsec:medium_relay}

A circular 1000-km orbit has period $2.46\,\mathrm{h}$, speed $3.12\,\mathrm{km\,s^{-1}}$, and $L_{\rm orb}=14.05\,\mu\mathrm{s}\,\mathrm{d}^{-1}$.  It is slower than $T_M$ by $1.90\,\mu\mathrm{s}\,\mathrm{d}^{-1}$.  The $J_2$ timing line is about $67\,\mathrm{ps}$ and degree-2 tesseral terms remain at the many-ps level.  Figure~\ref{fig:gmm3_degree_spectrum} shows that degrees in the tens remain relevant in the RMS rate screen, but no finite-arc minimum degree is established here; the retained field must be verified against the degree/order-120 reference reconstruction for the selected trajectory and observable.

\subsection{Areostationary and areosynchronous orbits}\label{subsec:aso}

For analytic budgeting, the monopole synchronous radius is
\begin{equation}
 r_{\rm ASO}=\Big(\frac{GM_{\mars}}{\Omega_{\mars}^2}\Big)^{1/3}=20427.7\,\mathrm{km},
\label{eq:auto:028}
\end{equation}
corresponding to altitude $17031.5\,\mathrm{km}$ above the equator.  This is the Keplerian monopole value; equilibrium longitude, station-keeping dynamics, and small radius corrections depend on the tesseral field and rotation model.  The speed is $1.448\,\mathrm{km\,s^{-1}}$ and
\begin{equation}
L_{\rm ASO}=3.02\,\mu\mathrm{s}\,\mathrm{d}^{-1}.
\label{eq:auto:029}
\end{equation}
An areostationary clock is therefore faster than $T_M$ by $9.13\,\mu\mathrm{s}\,\mathrm{d}^{-1}$.  Its rate relative to TT requires the common ephemeris term in Eq.~\eqref{eq:tau_tt_master}.

For an exactly equatorial spacecraft stationary in the Mars body-fixed frame, the static harmonics give constant clock-rate offsets rather than orbital-frequency timing lines. The $J_2$ offset is longitude independent; the tesseral offsets depend on the fixed longitude.  The degree-2 static gravity scale is still above threshold: the $J_2$ fractional scale is $\sim1.3\times10^{-15}$, and the $C_{22}/S_{22}$ longitude-dependent scale is typically $10^{-16}$--$10^{-15}$.  Inclination, eccentricity, and libration can introduce periodic signatures; pure longitude libration at fixed radius and latitude changes tesseral terms but not $J_2$. The $J_2$ line scale from Eq.~\eqref{eq:J2_line} is about $6.7\,\mathrm{ps}$ if the orbit samples latitude periodically.

The solar tide is mandatory at ASO.  For an areostationary clock fixed in the Mars body frame, the solar quadrupole is sampled at the Mars solar-day, or synodic, rate rather than the sidereal spin rate,
\begin{equation}
\Omega_{\rm syn}\equiv\Omega_{\mars}-n_{\mars\odot},\qquad
n_{\mars\odot}\simeq\sqrt{\frac{GM_\odot}{a_{\mars}^3}} .
\label{eq:mars_synodic_rate}
\end{equation}
In the coplanar, uniformly rotating approximation, the integrated quadrupole line is
\begin{equation}
P_{\odot,2}^{\rm ASO}\simeq -\frac{3}{8}\frac{GM_\odot r_{\rm ASO}^2}{c^2R_{\mars\odot}^3\Omega_{\rm syn}}\sin 2\psi_\odot,
\label{eq:auto:030}
\end{equation}
Here $R_{\mars\odot}$ is the Mars--Sun distance, treated as constant during the representative cycle. The signed phase is $\psi_\odot=\lambda_\odot^{\rm bf}-\lambda_{\rm ASO}$, where both longitudes are east positive, $\lambda_\odot^{\rm bf}$ is the longitude of the equatorial projection of the Mars--Sun direction, and $\lambda_{\rm ASO}$ is the fixed spacecraft longitude. Thus $\psi_\odot(T)=\psi_{\odot,0}-\Omega_{\rm syn}(T-T_0)$ in this approximation; $\psi_{\odot,0}$ fixes the epoch geometry. Integrating the positive $\cos2\psi_\odot$ rate term gives the negative sine in Eq.~\eqref{eq:auto:030}. A solar declination $\delta_\odot$ multiplies this twice-daily component by $\cos^2\delta_\odot$; changing distance, declination, and projected longitude require direct quadrature in a seasonal model.  The mean-distance amplitude is $\simeq0.28\,\mathrm{ps}$, equivalent to $0.083\,\mathrm{mm}$ one-way light range; at perihelion it is larger by Eq.~\eqref{eq:solar_tide_perihelion_factor}, $\simeq0.37\,\mathrm{ps}$.  Using $\Omega_{\rm syn}$ instead of $\Omega_{\mars}$ changes the amplitude by only about $0.15\%$, but it is the physically correct sampling frequency for a body-fixed areostationary clock.  This term is small compared with Mars gravity harmonics but exceeds the $0.1\,\mathrm{ps}$ threshold.

\subsection{Highly elliptical Mars relay orbit}
\label{subsec:heo}

As a representative highly elliptical orbit (HEO) for relay and navigation applications, we take
\begin{equation}
 h_p=300\,\mathrm{km},\qquad h_a=17000\,\mathrm{km}.
\label{eq:auto:031}
\end{equation}
Then
\begin{equation}
 a=12046.2\,\mathrm{km},\qquad e=0.6932,
\label{eq:auto:032}
\end{equation}
with period $11.15\,\mathrm{h}$.  The secular local term is
\begin{equation}
L_{\rm HEO}=5.13\,\mu\mathrm{s}\,\mathrm{d}^{-1},
\label{eq:auto:033}
\end{equation}
so the HEO clock is faster than $T_M$ by $7.02\,\mu\mathrm{s}\,\mathrm{d}^{-1}$.  No numerical TT rate is assigned without a consistent Mars Time Ephemeris.

The periodic clock signature is dominated by the kinematic-plus-monopole term.  The maximum Keplerian excursion is $(2GM_{\mars}/a c^2 n)e\simeq0.350\,\mu\mathrm{s}$, and Eq.~\eqref{eq:elliptic_series} gives the first Fourier amplitudes
\begin{equation}
A_1\simeq0.330\,\mu\mathrm{s},\qquad A_2\simeq0.103\,\mu\mathrm{s},\qquad A_3\simeq0.0478\,\mu\mathrm{s}.
\label{eq:auto:034}
\end{equation}
The corresponding one-way light-range amplitudes are $98.9\,\mathrm{m}$, $30.9\,\mathrm{m}$, and $14.3\,\mathrm{m}$.  These signatures are much larger than the gravity-harmonic ps-level lines and must be integrated directly along the osculating orbit.  Near periapsis, the Mars gravity field should still be evaluated with high degree and order because the spacecraft enters the LMO gravity environment.  Near apoapsis, the solar tide approaches the ASO scale and should be retained.  For the diagnostic HEO used in Sec.~\ref{sec:numerical_simulations}, the exact Sun/Phobos/Deimos external-tide half-amplitude is $0.101\,\mathrm{ps}$ at mean Mars--Sun distance.  Scaling the dominant solar part to Mars perihelion gives $\simeq0.136\,\mathrm{ps}$, so the term is not merely marginal; it is a retained above-threshold contribution over a perihelion-containing operational interval.

\subsection{Phobos- and Deimos-related orbits}\label{subsec:phobos_deimos_orbits}

A circular Mars-centered orbit near Phobos' mean orbital radius $a_P\simeq9376\,\mathrm{km}$ has period $7.66\,\mathrm{h}$, speed $2.14\,\mathrm{km\,s^{-1}}$, and
\begin{equation}
L_P=6.59\,\mu\mathrm{s}\,\mathrm{d}^{-1}.
\label{eq:auto:035}
\end{equation}
It is faster than $T_M$ by $5.56\,\mu\mathrm{s}\,\mathrm{d}^{-1}$ and has a Mars $J_2$ timing line of about $21.5\,\mathrm{ps}$.  The solar tide is only $0.018\,\mathrm{ps}$ for a circular orbit at this radius and is below the adopted timing threshold.  However, if the spacecraft performs close operations relative to Phobos, the Phobos self-potential must be included.  At $100\,\mathrm{km}$ from Phobos' center,
\begin{equation}
\frac{GM_P}{\rho c^2}\simeq7.9\times10^{-17}=6.8\,\mathrm{ps}\,\mathrm{d}^{-1},
\label{eq:auto:036}
\end{equation}
while at the Phobos surface it is about $7.1\times10^{-16}=61\,\mathrm{ps}\,\mathrm{d}^{-1}$.

A circular Deimos-distance orbit at $a_D\simeq23463\,\mathrm{km}$ has period $30.31\,\mathrm{h}$, speed $1.35\,\mathrm{km\,s^{-1}}$, and $L_D=2.63\,\mu\mathrm{s}\,\mathrm{d}^{-1}$.  It is faster than $T_M$ by $9.52\,\mu\mathrm{s}\,\mathrm{d}^{-1}$.  The Mars $J_2$ line is about $5.4\,\mathrm{ps}$, and the solar quadrupole timing line is about $0.45\,\mathrm{ps}$ at mean Mars--Sun distance and $0.60\,\mathrm{ps}$ at perihelion, both above the threshold.  Close Deimos operations require the Deimos self-potential; at $100\,\mathrm{km}$ from Deimos' center it contributes $1.1\times10^{-17}$, or $0.92\,\mathrm{ps}\,\mathrm{d}^{-1}$.  Tables~\ref{tab:orbit_summary} and~\ref{tab:range_equiv} consolidate the resulting secular clock rates, periodic timing signatures, and one-way range equivalents used in the simulation and implementation sections.

\begin{table*}[!tbp]
\caption{Representative Mars orbital clock rates and dominant periodic signatures. Numerical secular entries are leading monopole (or Keplerian kinetic-plus-monopole) estimates, not full-field rates.  For orbiters, ``Mean vs. $T_M$'' is $L_{\rm surf}^{\rm def}-L_{\rm orb}$; positive means the orbit clock is faster than $T_M$.  Rates relative to TT are not quoted because they require an ephemeris-specific $T_M-\TT$ product and, for a physical areoid anchor, Eq.~\eqref{eq:areoid_tt_bridge}.}
\begin{ruledtabular}
\setlength{\tabcolsep}{2.0pt}
\begin{tabular}{lccccc}
Regime & Radius/altitude & Period & $L_{\rm clk}$ & Mean vs. $T_M$ & Dominant local periodic terms\\
 & & & $(\mu\mathrm{s}/\mathrm{d})$ & $(\mu\mathrm{s}/\mathrm{d})$ & one-way amplitude\\
\hline
$T_M$ defining level & $W=W_{\rm def}$ & --- & 12.15091 & 0 & coordinate convention; not a physical clock\\
LMO circular & $h=300\,\mathrm{km}$ & 1.895 h & 16.71 & $-4.56$ & $J_2$: $86.7\,\mathrm{ps}$; tesseral: tens ps\\
LMO circular & $h=400\,\mathrm{km}$ & 1.973 h & 16.27 & $-4.12$ & $J_2$: $83.3\,\mathrm{ps}$; high-degree field retained\\
Relay circular & $h=1000\,\mathrm{km}$ & 2.458 h & 14.05 & $-1.90$ & $J_2$: $66.9\,\mathrm{ps}$\\
Phobos-distance & $r=9376\,\mathrm{km}$ & 7.657 h & 6.59 & $+5.56$ & $J_2$: $21.5\,\mathrm{ps}$; Phobos close potential\\
Areostationary & $r=20427.7\,\mathrm{km}$ & 24.623 h & 3.02 & $+9.13$ & Sun: $0.28/0.37\,\mathrm{ps}$; static $J_2$ offset\\
Deimos-distance & $r=23463\,\mathrm{km}$ & 30.310 h & 2.63 & $+9.52$ & $J_2$: $5.4\,\mathrm{ps}$; Sun: $0.45/0.60\,\mathrm{ps}$\\
HEO relay & $300\times17000\,\mathrm{km}$ & 11.150 h & 5.13 & $+7.02$ & $K+M$: $0.330,0.103,0.0478\,\mu\mathrm{s}$\\
\end{tabular}
\end{ruledtabular}
\label{tab:orbit_summary}
\end{table*}

\begin{table*}[!tbp]
\caption{Conversion of selected timing signatures into one-way light-range equivalents.  Solar-tide entries are quoted at perihelion when used for retention decisions.  The published Mars--Earth entries are physical-clock comparison quantities from Ref.~\cite{AshbyPatla2025}, not values of $T_M-\TT$ unless Eq.~\eqref{eq:areoid_tt_bridge} has been applied.}
\begin{ruledtabular}
\begin{tabular}{lcc}
Timing or rate quantity & Time amplitude & Equivalent one-way range\\
\hline
Formal timing threshold & $0.1\,\mathrm{ps}$ & $0.030\,\mathrm{mm}$\\
Fractional threshold over one day & $5\times10^{-18}\times86400\,\mathrm{s}=0.432\,\mathrm{ps}$ & $0.129\,\mathrm{mm}$\\
Published Mars-areoid mean vs. TT & $477.60\,\mu\mathrm{s}/\mathrm{d}$ & $143.2\,\mathrm{km}/\mathrm{d}$\\
Published Mars--Earth annual rate amplitude & $226.8\,\mu\mathrm{s}/\mathrm{d}$ & $68.0\,\mathrm{km}/\mathrm{d}$\\
BCRS--MCRS position term, surface & $0.91\,\mu\mathrm{s}$ & $273\,\mathrm{m}$\\
BCRS--MCRS position term, ASO & $5.48\,\mu\mathrm{s}$ & $1644\,\mathrm{m}$\\
Mars Sagnac, surface--300 km LMO (max.) & $9.9\,\mathrm{ns}$ & $3.0\,\mathrm{m}$\\
Mars Sagnac, surface--ASO (max.) & $55\,\mathrm{ns}$ & $16\,\mathrm{m}$\\
300-km LMO Mars $J_2$ line & $86.7\,\mathrm{ps}$ & $26.0\,\mathrm{mm}$\\
Areostationary solar-tide line, perihelion & $0.37\,\mathrm{ps}$ & $0.11\,\mathrm{mm}$\\
Deimos-distance solar-tide line, perihelion & $0.60\,\mathrm{ps}$ & $0.18\,\mathrm{mm}$\\
HEO first Fourier line & $0.330\,\mu\mathrm{s}$ & $98.9\,\mathrm{m}$\\
\end{tabular}
\end{ruledtabular}
\label{tab:range_equiv}
\end{table*}

The model hierarchy follows directly from the two retention gates: an instantaneous fractional-rate term larger than $5\times10^{-18}$ can bias frequency and potential products, while a zero-mean term with integrated one-way amplitude larger than $0.1\,\mathrm{ps}$ can bias synchronization and ranging even when its mean rate vanishes.  Table~\ref{tab:term_hierarchy} consolidates the dominant, regime-dependent, and sub-threshold terms used in the remainder of the paper.  It also indicates where a quantity is a defining convention rather than a measured geophysical result.

\begin{table*}[!tbp]
\centering
\caption{Dominant-to-negligible hierarchy of Mars clock and link terms for the adopted $5\times10^{-18}$ fractional-frequency and $0.1\,\mathrm{ps}$ one-way timing gates.  Values are representative scales from the equations and tables in the text; operational realizations must recompute them with the adopted ephemeris, gravity field, rotation model, and trajectory.}
\label{tab:term_hierarchy}
\renewcommand{\arraystretch}{1.08}
\begin{tabular}{llll}
\hline\hline
\parbox[t]{0.20\textwidth}{\raggedright Effect or correction} &
\parbox[t]{0.28\textwidth}{\raggedright Representative scale} &
\parbox[t]{0.25\textwidth}{\raggedright Retention decision} &
\parbox[t]{0.17\textwidth}{\raggedright Traceability} \\
\hline
\parbox[t]{0.20\textwidth}{\raggedright BCRS--MCRS origin rate and Mars annual term} &
\parbox[t]{0.28\textwidth}{\raggedright Sun-only mean $L_A^{(\odot)}\simeq0.840\,\mathrm{ms\,d^{-1}}$; Keplerian annual $P_A$ amplitude $\sim11\,\mathrm{ms}$.} &
\parbox[t]{0.25\textwidth}{\raggedright Mandatory; must be integrated into a Mars time ephemeris rather than approximated by a constant rate.} &
\parbox[t]{0.17\textwidth}{\raggedright Eqs.~\eqref{eq:autoalign:001}--\eqref{eq:mars_time_ephemeris_split}.} \\[0.8ex]
\parbox[t]{0.20\textwidth}{\raggedright Position-dependent BCRS--MCRS term} &
\parbox[t]{0.28\textwidth}{\raggedright $(\bvec v_{\mars}\cdot\bvec r_{\mars})/c^2$: $0.91\,\mu\mathrm{s}$ at the surface and $5.48\,\mu\mathrm{s}$ at ASO.} &
\parbox[t]{0.25\textwidth}{\raggedright Mandatory event-dependent term for transformations between BCRS and MCRS.} &
\parbox[t]{0.17\textwidth}{\raggedright Sec.~\ref{subsec:bcrs_mcrs}; Table~\ref{tab:range_equiv}.} \\[0.8ex]
\parbox[t]{0.20\textwidth}{\raggedright Mars surface scaling} &
\parbox[t]{0.28\textwidth}{\raggedright $L_{\rm surf}^{\rm def}=12.15091\,\mu\mathrm{s\,d^{-1}}$.} &
\parbox[t]{0.25\textwidth}{\raggedright Defining convention for $T_M$; site/areoid corrections are not part of the global scale.} &
\parbox[t]{0.17\textwidth}{\raggedright Sec.~\ref{subsec:mars_surface_scale}.} \\[0.8ex]
\parbox[t]{0.20\textwidth}{\raggedright Surface realization: convention offset, height, areoid, seasonal gravity} &
\parbox[t]{0.28\textwidth}{\raggedright $3.57\,\mathrm{ps\,d^{-1}}$ per m; $0.028\,\mathrm{m}$ gives $0.1\,\mathrm{ps}$ in one day; seasonal low-degree terms can reach $10^{-18}$--$10^{-17}$.} &
\parbox[t]{0.25\textwidth}{\raggedright Mandatory for landed clocks or carried explicitly as systematic error.} &
\parbox[t]{0.17\textwidth}{\raggedright Secs.~\ref{subsec:surface_site_offsets}, \ref{subsec:seasonal_tides}; Table~\ref{tab:realization_error_budget}.} \\[0.8ex]
\parbox[t]{0.20\textwidth}{\raggedright Circular-orbit secular rates} &
\parbox[t]{0.28\textwidth}{\raggedright $L_{\rm orb}=16.71\,\mu\mathrm{s\,d^{-1}}$ at 300 km LMO and $3.02\,\mu\mathrm{s\,d^{-1}}$ at ASO.} &
\parbox[t]{0.25\textwidth}{\raggedright Mandatory for every orbiter; sign relative to $T_M$ changes with altitude.} &
\parbox[t]{0.17\textwidth}{\raggedright Eq.~\eqref{eq:L_circ}; Table~\ref{tab:orbit_summary}.} \\[0.8ex]
\parbox[t]{0.20\textwidth}{\raggedright HEO kinematic+monopole term} &
\parbox[t]{0.28\textwidth}{\raggedright Maximum Keplerian excursion $0.350\,\mu\mathrm{s}$; Fourier amplitudes $0.330$, $0.103$, and $0.0478\,\mu\mathrm{s}$.} &
\parbox[t]{0.25\textwidth}{\raggedright Mandatory; use exact Keplerian expression or direct quadrature, not a low-order eccentricity approximation.} &
\parbox[t]{0.17\textwidth}{\raggedright Eqs.~\eqref{eq:elliptic_exact}--\eqref{eq:elliptic_series}; Fig.~\ref{fig:heo_spectrum_full}.} \\[0.8ex]
\parbox[t]{0.20\textwidth}{\raggedright Mars static gravity harmonics} &
\parbox[t]{0.28\textwidth}{\raggedright $J_2$ timing line: $86.7\,\mathrm{ps}$ at 300 km, $6.7\,\mathrm{ps}$ at ASO radius for polar sampling, $5.4\,\mathrm{ps}$ at Deimos distance; higher degrees control residuals.} &
\parbox[t]{0.25\textwidth}{\raggedright Mandatory in LMO and retained by residual test elsewhere; degree/order 120 is the diagnostic reference reconstruction, not geophysical truth.} &
\parbox[t]{0.17\textwidth}{\raggedright Eq.~\eqref{eq:J2_line}; Figs.~\ref{fig:gmm3_degree_spectrum}, \ref{fig:validation_full}.} \\[0.8ex]
\parbox[t]{0.20\textwidth}{\raggedright External tides and moon self-potentials} &
\parbox[t]{0.28\textwidth}{\raggedright Solar tide $<10^{-3}\,\mathrm{ps}$ in LMO; high-relay mean/perihelion amplitudes are $0.28/0.37\,\mathrm{ps}$ at ASO and $0.45/0.60\,\mathrm{ps}$ at Deimos distance; the HEO diagnostic is $0.101/0.136\,\mathrm{ps}$.  Phobos self-potential $6.8\,\mathrm{ps\,d^{-1}}$ at $100\,\mathrm{km}$.} &
\parbox[t]{0.25\textwidth}{\raggedright Solar tide retained for ASO/HEO/Deimos-distance clocks; Phobos/Deimos self-potentials retained for close moon operations.} &
\parbox[t]{0.17\textwidth}{\raggedright Eqs.~\eqref{eq:tide}, \eqref{eq:exact_tide_model}; Tables~\ref{tab:orbit_summary}, \ref{tab:simulation_validation}.} \\[0.8ex]
\parbox[t]{0.20\textwidth}{\raggedright Mars Sagnac and Mars Shapiro link terms} &
\parbox[t]{0.28\textwidth}{\raggedright Surface--LMO Sagnac near $10\,\mathrm{ns}$; surface--ASO Sagnac $55\,\mathrm{ns}$; Mars Shapiro coefficient $3.18\,\mathrm{ps}$ and grazing geometries tens of ps.} &
\parbox[t]{0.25\textwidth}{\raggedright Mandatory for precise link observables; not part of the local clock-only budget.} &
\parbox[t]{0.17\textwidth}{\raggedright Sec.~\ref{sec:time_transfer}; Table~\ref{tab:range_equiv}; Fig.~\ref{fig:operational_constraints}.} \\[0.8ex]
\parbox[t]{0.20\textwidth}{\raggedright Local $c^{-4}$ proper-time terms and distant perturbers} &
\parbox[t]{0.28\textwidth}{\raggedright Stationary-model algebraic $c^{-4}$ terms $\le6\times10^{-20}$; reported Jupiter tide $<3.6\times10^{-21}$ through HEO apoapsis.} &
\parbox[t]{0.25\textwidth}{\raggedright Specified stationary-model terms are below the gates on the stated arcs; a complete local metric and mission error budget require separate evaluation.} &
\parbox[t]{0.17\textwidth}{\raggedright Eqs.~\eqref{eq:c4_local_bound_main}, \eqref{eq:c4_sim_bound}; Table~\ref{tab:simulation_validation}.} \\
\hline\hline
\end{tabular}
\end{table*}

\section{Numerical simulations and model-retention diagnostics}\label{sec:numerical_simulations}

The preceding sections give analytic transformations and orbit-dependent scaling laws. The simulations in this section are intended to make those scaling laws traceable, to test model truncation quantitatively, and to document the model hierarchy and numerical checks used in the figures. The adopted target follows the cislunar template methodology and is used uniformly throughout the paper: retain every modeled contribution whose fractional-rate effect exceeds $5\times10^{-18}$ or whose one-way accumulated timing amplitude exceeds $0.1\,\mathrm{ps}$ \cite{Turyshev2026Cislunar}. The simulations are deterministic and noise free. They test the relativistic clock model itself; stochastic orbit-determination errors, tracking noise, atmospheric drag, maneuver errors, and estimator covariance must be added in a mission-specific implementation.

\subsection{Full deterministic model used for the figures}\label{subsec:full_deterministic_model}

For all Mars-local clock simulations the plotted rate integrand is
\begin{align}
\dot q_{\rm sim}(T) &= \frac{1}{c^2}\Big\{\frac{1}{2}V^2(T)+U_{0}(X)+U_{\rm GMM3}^{2:120}(T,\bvec X)+U_{\rm tide}^{\odot P D}(T,\bvec X)\Big\},\label{eq:full_sim_rate}\\
\frac{\dd\tau}{\dd\TCA} &=1-\dot q_{\rm sim}+\delta_{c^{-4}},\label{eq:full_sim_tau}
\end{align}
where $U_0=GM_{\mars}/X$ and $U_{\rm GMM3}^{2:120}$ is the complete static GMM-3 spherical-harmonic potential through degree and order 120. The GMM-3 file used in the simulation is the NASA/PDS SHADR product \texttt{gmm3\_120\_sha.tab}, with reference radius $R_0=3396.0\,\mathrm{km}$, $GM_{\mars}=4.2828372854187757\times10^{13}\,\mathrm{m^3\,s^{-2}}$ as in Table~\ref{tab:constants}, normalized coefficients, and maximum degree/order 120 \cite{Genova2016,Konopliv2016,PGDAGMM3}. The full harmonic potential evaluated in the rotating Mars body-fixed frame is
\begin{equation}
U_{\rm GMM3}^{2:120}=\frac{GM_{\mars}}{r}
\sum_{\ell=2}^{120}\sum_{m=0}^{\ell}\Big(\frac{R_0}{r}\Big)^{\ell}\bar P_{\ell m}(\sin\phi)
\Big[\bar C_{\ell m}\cos m\lambda+\bar S_{\ell m}\sin m\lambda\Big].
\label{eq:gmm3_full_model}
\end{equation}
The reported reconstruction uses the GMM-3 normalized coefficients and matching fully normalized associated Legendre functions. The coefficient normalization and reference-frame conventions must be preserved; the limits of the archived numerical provenance are stated in Sec.~\ref{subsec:independent_checks}.

The external-body part is not truncated as a Legendre series. For the Sun, Phobos, and Deimos it is evaluated as the exact point-mass tidal potential about the Mars origin,
\begin{equation}
U_{\rm tide}^{B}(T,\bvec X)=GM_B\bigg[\frac{1}{|\bvec R_B(T)-\bvec X|}-\frac{1}{R_B(T)}-\frac{\bvec R_B(T)\cdot\bvec X}{R_B^3(T)}\bigg],
\qquad B\in\{\odot,P,D\}.
\label{eq:exact_tide_model}
\end{equation}
This expression includes all tidal multipoles of those point masses and avoids a misleading low-order tide truncation near Phobos- and Deimos-distance trajectories. The deterministic demonstration uses circular Phobos and Deimos reference orbits with $a_P=9376\,\mathrm{km}$, $GM_P=7.087\times10^5\,\mathrm{m^3\,s^{-2}}$, $a_D=23463\,\mathrm{km}$, and $GM_D=9.620\times10^4\,\mathrm{m^3\,s^{-2}}$, consistent with the JPL satellite-parameter table used here \cite{JPLSSD}. The Sun is placed at $a_{\mars}=1.523679\,\mathrm{au}$ for the diagnostic figures, with the slow Mars orbital phase included; perihelion retain/discard decisions multiply the solar-tide contribution by $f_{\rm peri}$ from Eq.~\eqref{eq:solar_tide_perihelion_factor}. No solar, Phobos, or Deimos tidal multipole is truncated inside Eq.~\eqref{eq:exact_tide_model}; the full point-mass expression is evaluated directly. Jupiter is not included in the plotted local-clock curves because a conservative closest-approach bound on its Mars-centered tidal potential gives $|U_{\rm tide}^{J}|/c^2<3.6\times10^{-21}$ even at HEO apoapsis, more than three orders of magnitude below the rate target and less than $3.2\times10^{-4}\,\mathrm{ps/day}$. Finite-size multipoles of the Sun and the two moons are likewise below the diagnostic threshold for the plotted geometries, except for close moon-proximity operations where a moon-centered local gravity model would be required.

For a time series, the mean is removed and the one-way timing residual is reconstructed as
\begin{equation}
L_{\rm sim}=\langle\dot q_{\rm sim}\rangle,
\qquad
\delta\dot q_{\rm sim}(T)=\dot q_{\rm sim}(T)-L_{\rm sim},
\qquad
P_{\rm sim}(T)=\int_{T_0}^{T}\delta\dot q_{\rm sim}(T')\,\dd T'.
\label{eq:full_sim_reconstruction}
\end{equation}
The reported timing half-amplitudes use Eq.~\eqref{eq:retention_timing_metric}; for plotting, the arbitrary integration constant is shifted so that $\max P_{\rm sim}+\min P_{\rm sim}=0$.  This mid-range centering changes neither the half peak-to-peak amplitude nor any nonzero-frequency Fourier coefficient.
The integral is evaluated with the composite trapezoidal rule on uniform TCA grids. The high-eccentricity orbit uses a Newton solution of Kepler's equation with stopping tolerance $10^{-13}\,\mathrm{rad}$. The stated diagnostic procedure uses no smoothing or fitted clock signal; independent checks supplied with this revision are distinguished from the archived figures in Sec.~\ref{subsec:independent_checks}.

For the stationary monopole-plus-spin check, the algebraic terms omitted from Eq.~\eqref{eq:full_sim_rate} satisfy
\begin{equation}
 \beta_{\rm alg}=\frac{1}{c^4}\max_{T\in\mathcal I}
 \left\{\frac18V^4+\frac32V^2|U|+\frac12U^2+4|W^i_{\mars}V_i|\right\},
 \label{eq:c4_sim_bound}
\end{equation}
with $U=GM_{\mars}/r$ in this check and the spin bound specified after Eq.~\eqref{eq:c4_local_bound_main}. The resulting conservative bounds are $4\times10^{-20}$ for the 300 km LMO and $6\times10^{-20}$ for the HEO, below the rate target by factors of 125 and 83, respectively. Their accumulated bounds over the stated arcs are also below the timing gate. These are bounds on that model's algebraic terms, not on the complete post-Newtonian local scalar potential, gravity uncertainty, or mission realization. Tables~\ref{tab:full_simulation_setup} and~\ref{tab:simulation_geometry} summarize the diagnostic inputs, and Table~\ref{tab:simulation_validation} records the scope of the residual checks.

\begin{table*}[!tbp]
\centering
\caption{Traceable simulation inputs and retained deterministic physics with overall conditions summarized in Table~\ref{tab:simulation_geometry}. The threshold policy follows the lunar/cislunar high-precision time-scale methodology of Refs.~\cite{Turyshev2025Time,Turyshev2026Cislunar}.}
\label{tab:full_simulation_setup}
\renewcommand{\arraystretch}{1.05}
\begin{tabular}{ll}
\hline\hline
Item & Traceable specification \\
\hline
\parbox[t]{0.25\linewidth}{\raggedright Target accuracy} & \parbox[t]{0.70\linewidth}{\raggedright Retain terms with fractional-rate amplitude $>5\times10^{-18}$ or one-way accumulated timing amplitude $>0.1\,\mathrm{ps}$.}\\[0.7ex]
\parbox[t]{0.25\linewidth}{\raggedright Mars static gravity included} & \parbox[t]{0.70\linewidth}{\raggedright Complete GMM-3 normalized static field through degree/order 120, evaluated from the NASA/PDS SHADR coefficient file named \texttt{gmm3\_120\_sha.tab}; see Eq.~\eqref{eq:gmm3_full_model}.}\\[0.7ex]
\parbox[t]{0.25\linewidth}{\raggedright External-body terms included} & \parbox[t]{0.70\linewidth}{\raggedright Exact point-mass tidal potentials for the Sun, Phobos, and Deimos, with the origin and dipole terms removed; see Eq.~\eqref{eq:exact_tide_model}.}\\[0.7ex]
\parbox[t]{0.25\linewidth}{\raggedright Spacecraft trajectories} & \parbox[t]{0.70\linewidth}{\raggedright Exact circular Mars orbits for altitude sweeps; a 300 km near-polar LMO; and a representative HEO with $h_p=300\,\mathrm{km}$, $h_a=17000\,\mathrm{km}$, $a=12046.19\,\mathrm{km}$, $e=0.693165225$, and $T=11.1502895\,\mathrm{h}$.}\\[0.7ex]
\parbox[t]{0.25\linewidth}{\raggedright Numerical reconstruction} & \parbox[t]{0.70\linewidth}{\raggedright Mean removal using Eq.~\eqref{eq:full_sim_reconstruction}, composite-trapezoidal quadrature, mid-range centering of the arbitrary integration constant, and Fourier projection of the same time-domain residual.}\\
\hline\hline
\end{tabular}
\end{table*}

\begin{table*}[t]
\caption{Geometric and numerical assumptions for the deterministic
simulation figures.  These assumptions define the diagnostic calculation;
they are not a substitute for mission orbit determination.}
\label{tab:simulation_geometry}
\begin{ruledtabular}
\begin{tabular}{ll}
\parbox[t]{0.18\textwidth}{\raggedright Quantity} &
\parbox[t]{0.76\textwidth}{\raggedright Adopted diagnostic value or convention} \\
\hline
\parbox[t]{0.18\textwidth}{\raggedright Time argument} &
\parbox[t]{0.76\textwidth}{\raggedright \(T=\TCA\) for Mars-local integrations; ephemeris time arguments and scaling explicitly identified} \\[0.4ex]
\parbox[t]{0.18\textwidth}{\raggedright Reference epoch} &
\parbox[t]{0.76\textwidth}{\raggedright Diagnostic epoch only; complete absolute phases are not specified in the archived figure set (Sec.~\ref{subsec:independent_checks})} \\[0.4ex]
\parbox[t]{0.18\textwidth}{\raggedright Mars rotation} &
\parbox[t]{0.76\textwidth}{\raggedright Uniform rotation for the diagnostic body-fixed gravity reconstruction; not an operational orientation realization} \\[0.4ex]
\parbox[t]{0.18\textwidth}{\raggedright LMO geometry} &
\parbox[t]{0.76\textwidth}{\raggedright Circular, $i=92^\circ$, 300 km diagnostic; $T=0$ at the ascending node; Fig.~\ref{fig:lmo_full_timeseries} uses 4600 samples over three periods} \\[0.4ex]
\parbox[t]{0.18\textwidth}{\raggedright HEO geometry} &
\parbox[t]{0.76\textwidth}{\raggedright Keplerian ellipse with $h_p=300\,{\rm km}$, $h_a=17000\,{\rm km}$, $e=0.693165225$; $T=0$ at periapsis} \\[0.4ex]
\parbox[t]{0.18\textwidth}{\raggedright External bodies} &
\parbox[t]{0.76\textwidth}{\raggedright Sun at mean $a_{\mars}$ for figures; multiply the solar tide by $f_{\rm peri}=1.342$ for retention; Phobos/Deimos on circular diagnostic orbits} \\[0.4ex]
\parbox[t]{0.18\textwidth}{\raggedright Quadrature} &
\parbox[t]{0.76\textwidth}{\raggedright Composite trapezoidal rule after mean removal of \(\dot q_{\rm sim}\)} \\[0.4ex]
\parbox[t]{0.18\textwidth}{\raggedright Kepler solver} &
\parbox[t]{0.76\textwidth}{\raggedright Newton iteration with stopping tolerance \(10^{-13}\,{\rm rad}\)} \\[0.4ex]
\parbox[t]{0.18\textwidth}{\raggedright Validation} &
\parbox[t]{0.76\textwidth}{\raggedright Degree truncation against GMM-3 degree/order 120; step-size convergence; one-period closure} \\
\end{tabular}
\end{ruledtabular}
\end{table*}

\begin{table*}[!tbp]
\centering
\caption{Quantified omissions, deterministic residuals, and validation checks for the simulations summarized in Table~\ref{tab:full_simulation_setup}.  The deterministic calculations are reference calculations for model ranking and truncation control; a flight realization must replace the diagnostic trajectories with the mission orbit-determination solution.}
\label{tab:simulation_validation}
\renewcommand{\arraystretch}{1.05}
\begin{tabular}{ll}
\hline\hline
\parbox[t]{0.25\textwidth}{\raggedright Item} & \parbox[t]{0.70\textwidth}{\raggedright Traceable specification} \\
\hline

\parbox[t]{0.25\textwidth}{\raggedright Terms retained near the threshold} &
\parbox[t]{0.70\textwidth}{\raggedright Exact point-mass Sun, Phobos, and Deimos tides are included in the plotted clock model.  The diagnostic mean-distance one-way timing half-amplitudes are
\(4.98\times10^{-4}\,\mathrm{ps}\) for the 300 km LMO case and
\(0.101\,\mathrm{ps}\) for the representative HEO.  Applying the Mars perihelion multiplier of Eq.~\eqref{eq:solar_tide_perihelion_factor} gives \(6.68\times10^{-4}\,\mathrm{ps}\) and \(0.136\,\mathrm{ps}\), respectively, for the dominant solar part.  The HEO external tide is therefore a retained above-threshold deterministic contribution over perihelion-containing operational intervals, not an omitted residual.} \\[1.0ex]

\parbox[t]{0.25\textwidth}{\raggedright Terms omitted from plotted curves} &
\parbox[t]{0.70\textwidth}{\raggedright Local post-Newtonian terms (only the specified stationary-model algebraic part is bounded by Eq.~\eqref{eq:c4_sim_bound}); Jupiter's point-mass tide, bounded above by \(3.6\times10^{-21}\) in fractional rate through HEO apoapsis (larger radii require rescaling); finite-size multipoles of external bodies except during close moon-proximity operations; and stochastic force, measurement, and clock-noise terms, which are mission-specific.} \\[1.0ex]

\parbox[t]{0.25\textwidth}{\raggedright Residual deterministic error after the stated truncations} &
\parbox[t]{0.70\textwidth}{\raggedright The stationary-model algebraic bounds are $4\times10^{-20}$ for LMO and $6\times10^{-20}$ for HEO (factors 125 and 83 below the rate gate). Their unsubtracted time bounds are $0.00082$ ps over three LMO revolutions and $0.0025$ ps over one HEO revolution. These are not bounds on the full local potential or mission covariance.  Gravity-field truncation residuals are reported separately by differencing each truncated field against the degree/order-120 GMM-3 reference.}\\

\hline\hline
\end{tabular}
\end{table*}

\subsection{Gravity-field completeness and circular-orbit retention}\label{subsec:gravity_completeness}

\begin{figure}[ht]
\centering
\includegraphics[width=0.80\textwidth]{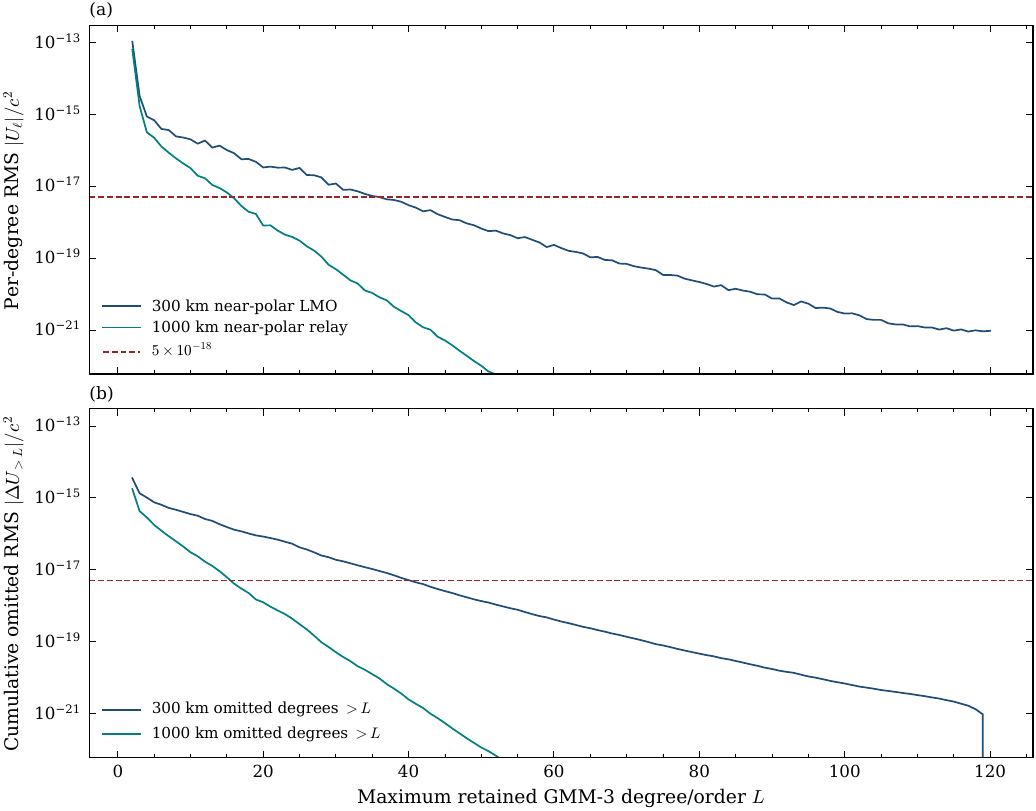}
\caption{GMM-3 static-gravity degree screening for representative low Mars orbits.  The upper panel gives the per-degree RMS fractional-rate contribution over one circular near-polar revolution; the lower panel gives the RMS cumulative omitted contribution after truncation at degree/order $L$.  The horizontal line compares these screening quantities with the nominal $5\times10^{-18}$ gate.  Because an RMS does not certify the maximum-rate criterion in Eq.~\eqref{eq:retention_rate_metric}, any retained degree must also be checked with the maximum-rate and finite-arc timing metrics.  The input is Eq.~\eqref{eq:gmm3_full_model} and the degree/order-120 GMM-3 SHADR product \cite{Genova2016,Konopliv2016,PGDAGMM3}.}
\label{fig:gmm3_degree_spectrum}
\end{figure}

Figure~\ref{fig:gmm3_degree_spectrum} ranks the static-gravity content by degree for representative low Mars orbits.  The plotted per-degree and cumulative quantities are RMS fractional rates, not the maximum residual in Eq.~\eqref{eq:retention_rate_metric}.  At 300 km the per-degree RMS remains above $5\times10^{-18}$ to about degree 40; at 1000 km the tail decays faster.  These curves are screening diagnostics.  A truncation is certified only after direct maximum-rate and finite-arc timing residuals are evaluated for the specified trajectory.

\begin{figure}[ht]
\centering
\includegraphics[width=0.80\textwidth]{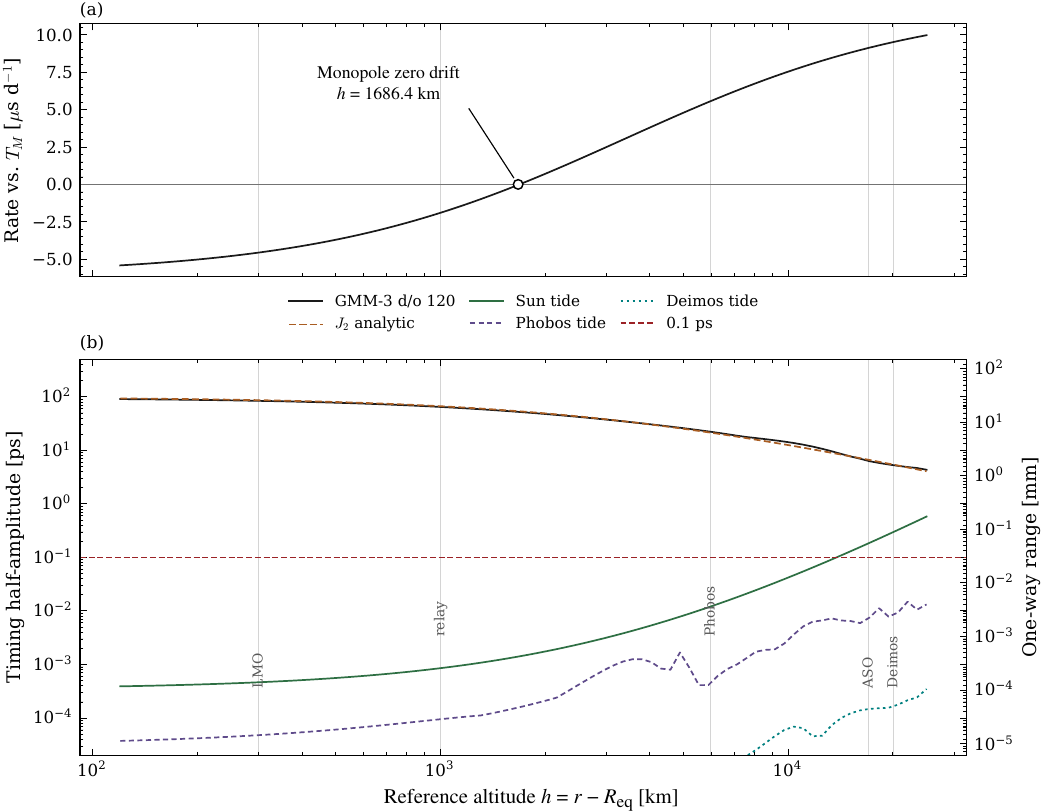}
\caption{Circular Mars-orbit clock budget. Panel (a) shows the leading monopole secular rate relative to $T_M$; the marked zero is $h_{\rm TA}=1686.4$ km from Eq.~\eqref{eq:time_aligned_radius}. Panel (b) retains the archived full GMM-3, analytic $J_2$, and exact point-mass Sun/Phobos/Deimos timing curves. These are finite-arc half-amplitudes on circular near-polar trajectories, after mean removal and integration over one spacecraft revolution, not instantaneous tides or phase-maximized envelopes. The horizontal coordinate $h=r-R_{\rm eq}$ is altitude relative to the equatorial reference radius, not an equatorial orbit designation; the ASO marker denotes radius only. Moon motion, relative phase, and the altitude-dependent integration interval permit non-monotonic curves, whereas the distant, slowly changing solar quadrupole gives a smooth curve. The Sun is at mean heliocentric distance; $f_{\rm peri}=1.342$ rescales its leading amplitude at fixed geometry, not the complete seasonal or phase envelope. The dashed line is the 0.1 ps timing gate. See Sec.~\ref{subsec:independent_checks} for independent checks and the original-data limits.}
\label{fig:circular_budget_full}
\end{figure}

Figure~\ref{fig:circular_budget_full} extends the orbit-by-orbit budget across practical circular radii. The lower panel compares the integrated timing half-amplitude of the full degree/order-120 gravity signal with the closed-form $J_2$ line and the exact point-mass tides from the Sun, Phobos, and Deimos. The near equality of the full GMM-3 and $J_2$ curves at the scale shown demonstrates that degree 2 dominates the absolute periodic amplitude, while Figs.~\ref{fig:gmm3_degree_spectrum} and~\ref{fig:validation_full} show that higher degrees still control sub-picosecond residuals. Figure~\ref{fig:gmm3_degree_spectrum} is an RMS screening diagnostic, whereas Fig.~\ref{fig:validation_full} applies the finite-arc timing metric to prescribed trajectories. The two diagnostics answer different questions, and neither alone establishes a universal minimum retained degree.

The altitude coordinate in Fig.~\ref{fig:circular_budget_full} is $h=r-R_{\rm eq}$; it does not make the plotted near-polar trajectories equatorial. The lower panel is the half peak-to-peak value of an \emph{integrated, mean-subtracted clock contribution over one spacecraft revolution}, not the instantaneous tidal potential. Its averaging interval has length $\Delta(r)=2\pi/n(r)$. Changing altitude therefore changes both the separation history and the interval over which a moving perturber is sampled.

For a nearly fixed distant Sun, the quadrupole rate amplitude is proportional to $r^2$ and integration supplies a factor of order $1/n$, giving the smooth representative scaling $A_{t,\odot}\propto r^{7/2}$. Smoothness versus altitude does not mean constancy in time. For a moon, $d_B^2=R_B^2+r^2-2R_Br\cos\chi_B(T)$ varies with the spacecraft and moon phases; $\chi_B$ is their instantaneous angular separation as seen from Mars. Both the changing geometry and combinations of their frequencies enter Eq.~\eqref{eq:exact_tide_model}. To exhibit the effect of finite-arc mean removal, a rate component $y(u)=C\cos(\omega u+\varphi)$ on $0\le u\le\Delta$ gives
\begin{equation}
 P_y(u)=\frac{C}{\omega}\left[\sin(\omega u+\varphi)-\sin\varphi
 -\frac{u}{\Delta}\{\sin(\omega\Delta+\varphi)-\sin\varphi\}\right].
 \label{eq:finite_window_response}
\end{equation}
Its amplitude depends on $\omega\Delta$ and phase, not only on $C$. The apparent $1/\omega$ singularity is removable as $\omega\to0$ because a constant rate is subtracted. Non-monotonic moon curves are therefore possible without a dynamical resonance; prescribed circular trajectories do not establish such a resonance. Individual features of an altitude sweep require phase and resolution tests. The independently specified checks in Sec.~\ref{subsec:independent_checks} demonstrate this behavior, but do not identify every ripple of the archived curves. No single-phase curve is a phase-maximized retention bound.

\subsection{Full time-domain simulation for low Mars orbit}\label{subsec:lmo_timeseries}

\begin{figure}[ht]
\centering
\includegraphics[width=0.80\textwidth]{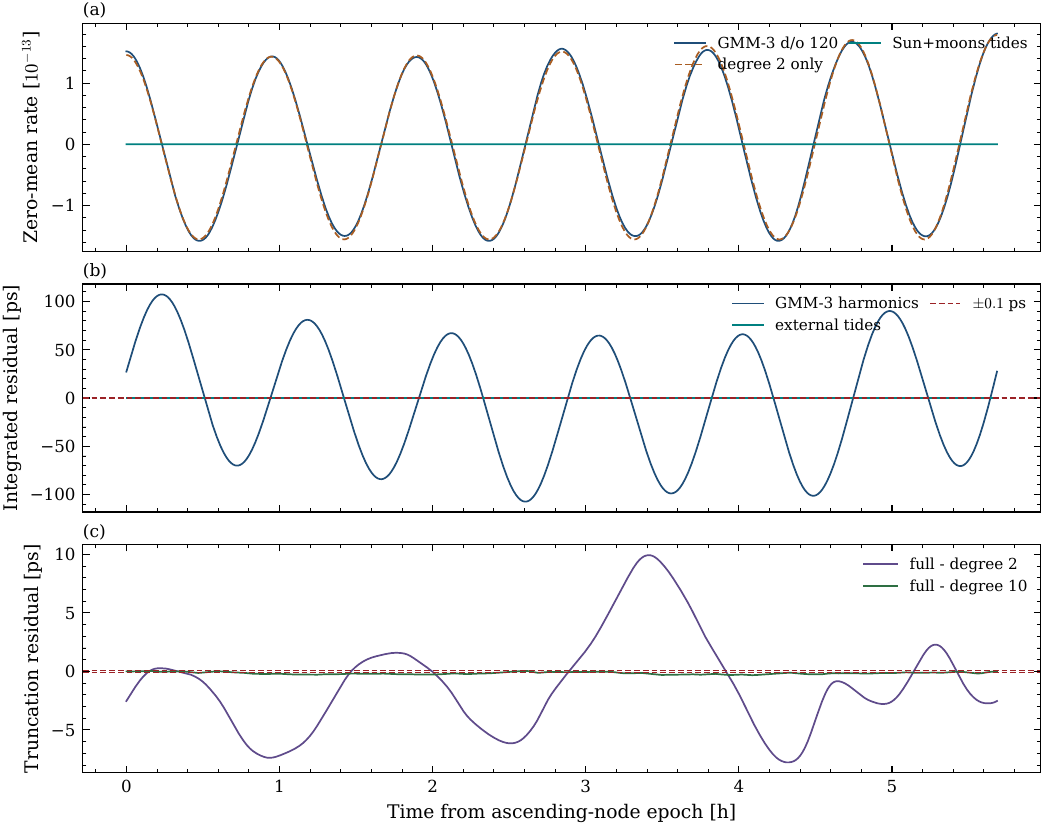}
\caption{Archived time-domain clock diagnostic for a 300 km near-polar low Mars orbit. The stated model is Eq.~\eqref{eq:full_sim_rate}, with GMM-3 degree/order 120 and exact Sun/Phobos/Deimos tides, sampled at 4600 epochs over three orbital periods. Panels (a), (b), and (c) show zero-mean rate, integrated timing, and direct full-minus-truncated residuals. The degree-2 and degree-10 curves in panel (c) have graph-read half-amplitudes of approximately 9 ps and 0.2 ps; the latter exceeds 0.1 ps. These are not the separate benchmark values in Table~\ref{tab:gmm3_truncation_selected}, and that table does not establish the minimum degree for this three-revolution arc.}
\label{fig:lmo_full_timeseries}
\end{figure}

Figure~\ref{fig:lmo_full_timeseries} shows the reported 300 km, $92^\circ$ near-polar circular diagnostic over three revolutions. Uniform Mars rotation changes the body-fixed longitude and samples the tesseral field. Panels (a) and (b) show zero-mean rate and integrated timing; panel (c) differences the full field against degree 2 and degree 10. The archived vector graphic gives approximate three-revolution timing half-amplitudes of $9$ ps and $0.2$ ps, respectively. These graph-read scales are not the $4.51$ ps and $0.111$ ps values of the separately reported truncation benchmark in Table~\ref{tab:gmm3_truncation_selected}. In particular, that benchmark cannot certify a degree-16 minimum for the three-revolution arc. The degree-10 residual in the displayed arc is above the timing gate; exact extrema and a revised minimum retained degree require the original arrays and matched averaging interval.

\subsection{Full time-domain simulation for a highly elliptical relay orbit}\label{subsec:heo_timeseries}

\begin{figure}[ht]
\centering
\includegraphics[width=0.80\textwidth]{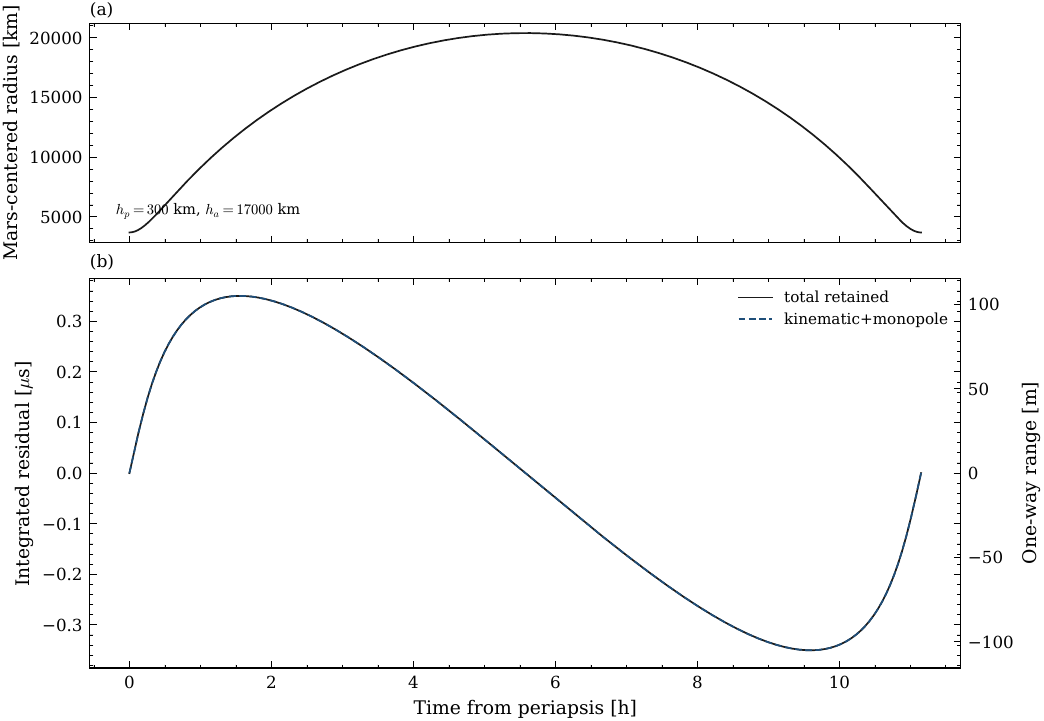}
\caption{Dominant timing behavior for the representative $300\,\mathrm{km}\times17000\,\mathrm{km}$ highly elliptical Mars relay orbit.  The state is an exact Keplerian orbit with $a=12046.19\,\mathrm{km}$, $e=0.693165225$, and $T=11.1502895\,\mathrm{h}$.  Panel (a) shows the Mars-centered radius; panel (b) shows the integrated residual from the full retained model and from the kinematic-plus-monopole component alone.  The near coincidence of the curves at the microsecond scale shows that secondary gravity and tide terms must be inspected separately, as in Fig.~\ref{fig:heo_secondary}.}
\label{fig:heo_primary}
\end{figure}

\begin{figure}[ht]
\centering
\includegraphics[width=0.80\textwidth]{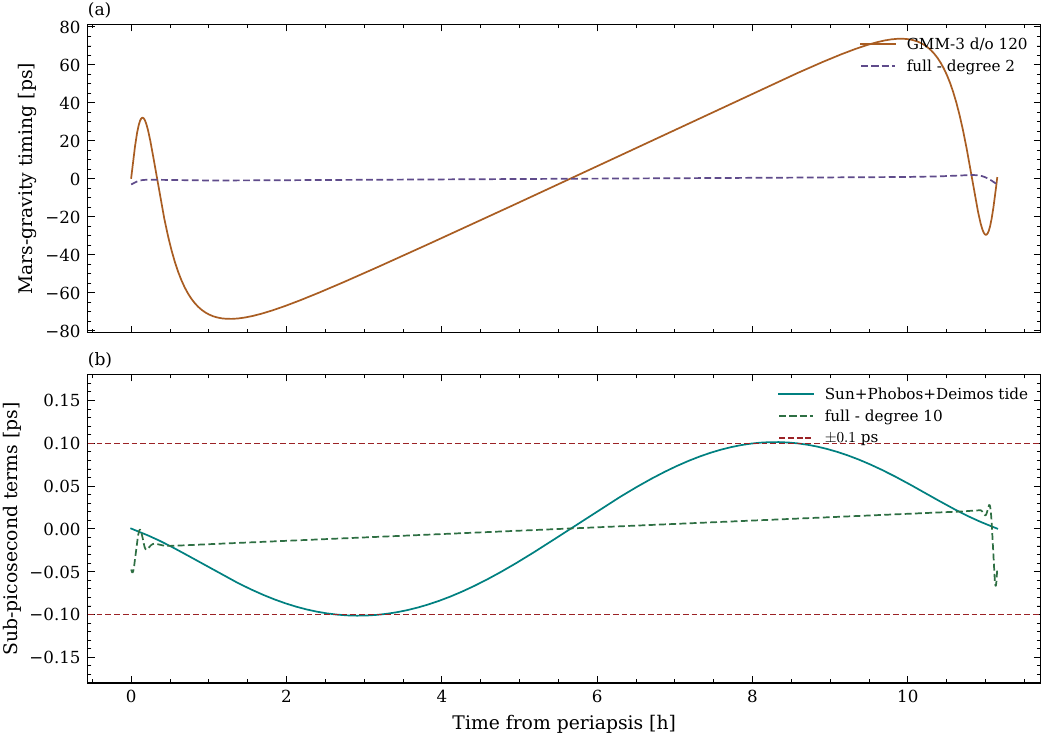}
\caption{Secondary timing terms for the same HEO as Fig.~\ref{fig:heo_primary}.  Panel (a) isolates the Mars static-gravity timing residual reconstructed from the full GMM-3 degree/order-120 field and its difference from the degree-2 reconstruction.  Panel (b) shows exact Sun/Phobos/Deimos tides and the direct full-minus-degree-10 truncation residual.  The dashed red lines mark the $\pm0.1\,\mathrm{ps}$ retention gate; the external-tide half-amplitude is $0.101\,\mathrm{ps}$ in the plotted mean-distance diagnostic and $\simeq0.136\,\mathrm{ps}$ after perihelion scaling of the dominant solar part, while the degree-10 residual is below the gate for this trajectory.}
\label{fig:heo_secondary}
\end{figure}

\begin{figure}[ht]
\centering
\includegraphics[width=0.70\textwidth]{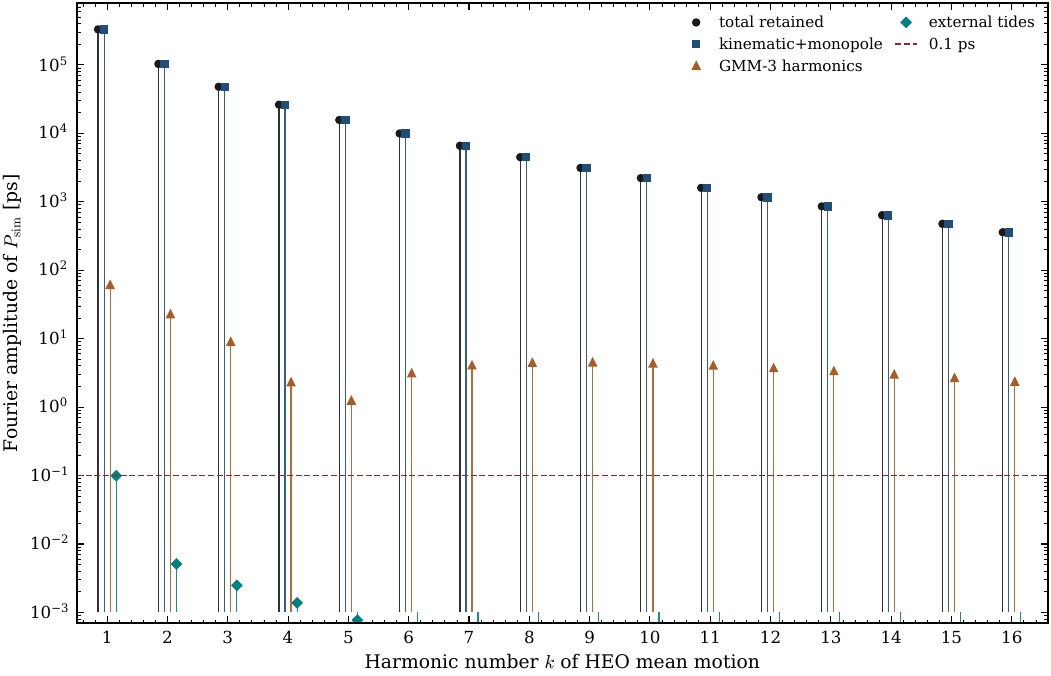}
\caption{Fourier amplitudes of the HEO timing residuals.  The plotted points are direct sine/cosine projections of the time-domain arrays used in Figs.~\ref{fig:heo_primary} and \ref{fig:heo_secondary}; no fitted analytic series is introduced.  The kinematic-plus-monopole term dominates many harmonics of the mean motion, while Mars harmonics remain above the $0.1\,\mathrm{ps}$ line over a broad range.  The figure demonstrates why low-order eccentricity expansions are inadequate for sub-picosecond timing in high-eccentricity relay orbits.}
\label{fig:heo_spectrum_full}
\end{figure}

High-eccentricity relay trajectories are the regimes in which closed-form low-order eccentricity expansions are most likely to fail.  We therefore split the HEO diagnostic into three figures with a clearer hierarchy.  Figure~\ref{fig:heo_primary} shows the orbital radius and the dominant integrated timing residual.  Figure~\ref{fig:heo_secondary} isolates the Mars-harmonic and sub-picosecond residuals that determine the retention rule.  Figure~\ref{fig:heo_spectrum_full} then projects the same time-domain arrays onto harmonics of the HEO mean motion.  All three figures use the same full deterministic model summarized in Tables~\ref{tab:full_simulation_setup} and~\ref{tab:simulation_validation}; no new assumptions are introduced between the time-domain and spectral products.

\subsection{Truncation and numerical convergence}\label{subsec:truncation_convergence}

\begin{figure}[ht]
\centering
\includegraphics[width=0.76\textwidth]{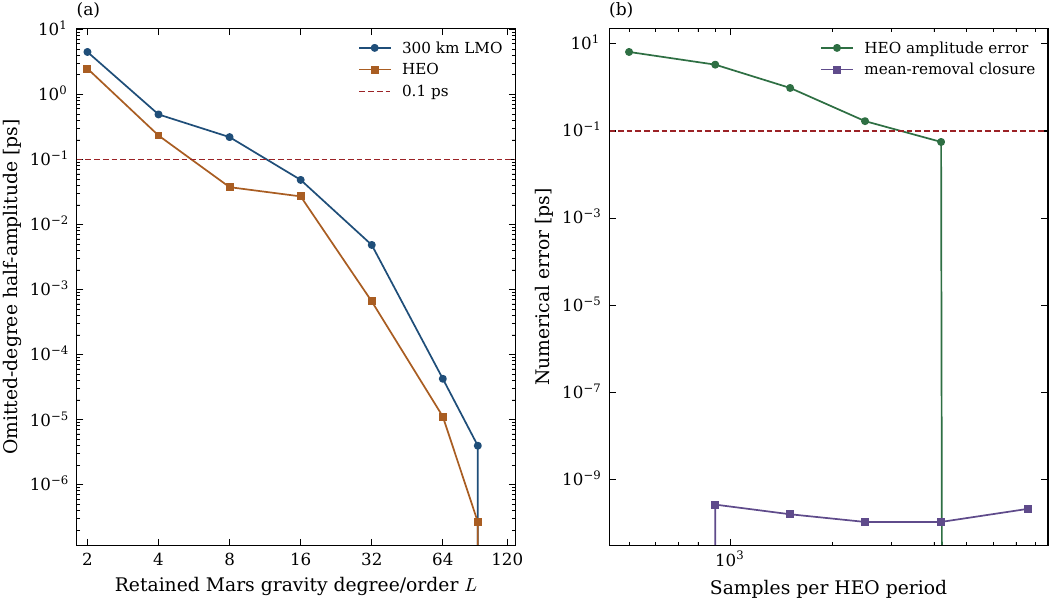}
\caption{Archived model-truncation and numerical-convergence diagnostic. Panel (a) displays the reported full-minus-truncated timing benchmark; its LMO curve is not the three-revolution arc in Fig.~\ref{fig:lmo_full_timeseries}. Panel (b) compares HEO quadrature with the densest grid used in the original calculation. The dashed line is 0.1 ps. Agreement with that grid and mean-removal closure do not alone establish an absolute integration-error bound. Independent exact-solution checks and the limits of the original numerical provenance are given in Sec.~\ref{subsec:independent_checks}.}
\label{fig:validation_full}
\end{figure}

Figure~\ref{fig:validation_full} and Table~\ref{tab:gmm3_truncation_selected} preserve the reported static-field truncation benchmark. The tabulated LMO residuals decrease from 4.51 ps at degree 2 to 0.111 ps at degree 10 and 0.0485 ps at degree 16; the HEO benchmark decreases from 2.50 ps at degree 2 to 0.0375 ps at degree 8. Their qualitative implication---higher degrees matter for sub-picosecond timing---is supported by the archived figures. However, the LMO benchmark is not the three-revolution residual displayed in Fig.~\ref{fig:lmo_full_timeseries}. Complete benchmark phase, interval, and sampling metadata are not available in the numerical archive, so these numbers are not used to certify a retained degree for another arc or a universal minimum degree.

Panel (b) compares HEO quadrature against the densest grid in that reported calculation. Mean-removal closure is an algebraic consistency check, not independent proof of integration accuracy. An independent comparison with the exact Keplerian clock solution, including pointwise rather than only amplitude error, is provided in Sec.~\ref{subsec:independent_checks}. It does not retroactively specify the original integration grid.

The degree/order-120 result is a reference reconstruction within one static gravity solution, not geophysical truth.  The differences shown here omit coefficient covariance, time-variable gravity, Mars-orientation uncertainty, and errors in $R_0$ or $GM_{\mars}$.  They are also evaluated on the same prescribed state history, so they isolate the direct potential contribution; a flight solution must additionally repropagate or re-estimate the trajectory and carry the joint orbit, gravity, orientation, and clock covariance.

\begin{table}[ht]
\centering
\caption{Reported GMM-3 truncation benchmark accompanying Fig.~\ref{fig:validation_full}. Values are timing half-amplitudes relative to the degree/order-120 reference. The LMO entries are not the three-revolution residuals of Fig.~\ref{fig:lmo_full_timeseries}. Missing original phase/arc metadata prevents their use as a retained-degree certification; they are preserved as reported diagnostic values. The degree-120 difference is zero by definition.}
\label{tab:gmm3_truncation_selected}
\begin{ruledtabular}
\begin{tabular}{lcc}
Retained degree/order $L$ &
300 km LMO residual & HEO residual \\
 & $(\mathrm{ps})$ & $(\mathrm{ps})$ \\
\hline
2  & $4.51$ & $2.50$ \\
4  & $\cdots$ & $0.234$ \\
8  & $0.221$ & $0.0375$ \\
10 & $0.111$ & $\cdots$ \\
16 & $0.0485$ & $\cdots$ \\
32 & $4.85\times10^{-3}$ & $\cdots$ \\
120 & $0$ & $0$ \\
\end{tabular}
\end{ruledtabular}
\end{table}

\subsection{Surface realization and link-level relativistic scales}\label{subsec:surface_link_scales}

\begin{figure}[ht]
\centering
\includegraphics[width=0.85\textwidth]{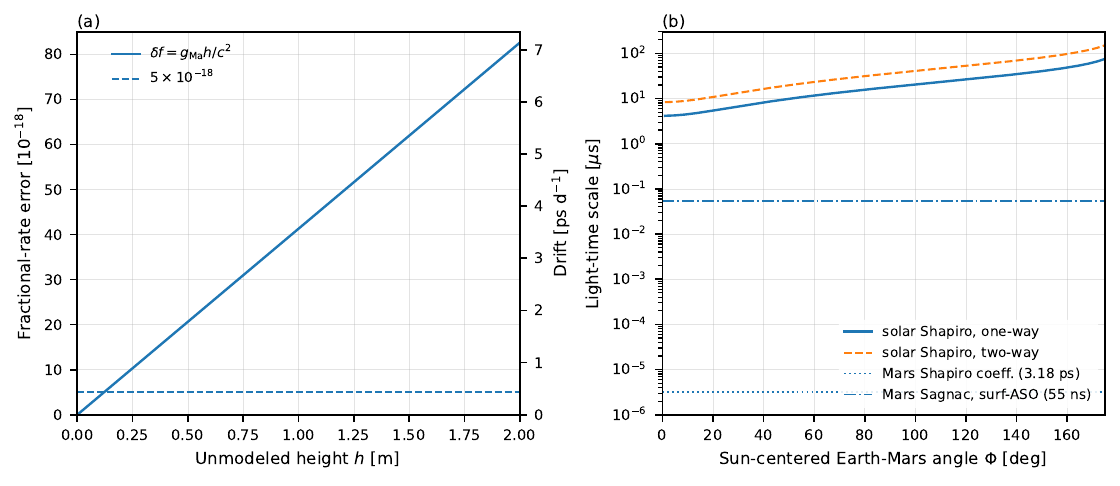}
\caption{Operational realization scales that accompany the formal clock model.  Panel (a) converts an unmodeled Mars site-height or areoid error into fractional frequency using $\delta f=g_{\mars}h/c^2$ and into one-day accumulated drift.  Panel (b) shows the one- and two-way solar Shapiro delay for a simplified Earth--Mars geometry, together with the Mars Shapiro coefficient and the first-order surface-to-areostationary Sagnac scale.  The Sagnac line is plotted at $5.5\times10^{-2}\,\mu\mathrm{s}=55\,\mathrm{ns}$, corresponding to a $16\,\mathrm{m}$ one-way light-time effect.  The plotted quantities are not stochastic measurement errors; they identify deterministic geodetic and link-relativity terms that must be modeled before the formal $0.1\,\mathrm{ps}$ clock accuracy can be realized.}
\label{fig:operational_constraints}
\end{figure}

The preceding figures test the local clock model.  Figure~\ref{fig:operational_constraints} connects that model to two operational limitations: geodetic realization of the Mars surface scale and relativistic link propagation.  Panel (a) shows that a height or areoid error of $0.121\,\mathrm{m}$ corresponds to $5\times10^{-18}$ in fractional frequency, while $0.028\,\mathrm{m}$ corresponds to $0.1\,\mathrm{ps}$ accumulated over one day.  Panel (b) shows that solar Shapiro delay for an Earth--Mars link is many microseconds to hundreds of microseconds depending on geometry (the plotted abscissa is the Sun-centered Earth--Mars angle, not the apparent solar elongation); the Mars Shapiro coefficient is picoseconds, while the surface-to-areostationary Sagnac scale is about $55\,\mathrm{ns}$, corresponding to $16\,\mathrm{m}$ of one-way light range.

\subsection{Interpretation for model retention}\label{subsec:model_retention_interpretation}

The diagnostics support a stricter retention policy than a leading-term analytic treatment. Low Mars orbit should use a high-degree Mars gravity field in the clock model, not only in the orbit propagator. The three-revolution 300 km near-polar plot exceeds the $0.1\,\mathrm{ps}$ timing goal at degree 10. The smaller degree-16 and degree-32 values in the separate archived benchmark do not certify that longer arc. A retained degree must be tested on the actual phase, interval, and observable. The degree/order-120 reference is used primarily for the fractional-rate screen and for reproducible differencing; the timing gate then quantifies how far a lower-degree approximation can be simplified for a specified arc and observable. Highly elliptical relays require direct numerical quadrature because the kinematic-plus-monopole residual is hundreds of nanoseconds, the full Mars-harmonic residual is tens of picoseconds, and the exact Sun/Phobos/Deimos tide is $0.101\,\mathrm{ps}$ in the mean-distance diagnostic but grows to $\simeq0.136\,\mathrm{ps}$ after perihelion scaling of the dominant solar part. High relay and areostationary regimes must retain the solar tide and Mars tesseral terms; Phobos and Deimos must be modeled with exact point-mass or moon-centered potentials for close operations. Finally, a Mars surface time scale is not limited primarily by the formal post-Newtonian transformation but by the realization of site potential, height, seasonal gravity, and station coordinates.

\subsection{Independent checks and numerical provenance}\label{subsec:independent_checks}

The accompanying \texttt{verify\_revision.py} supplies new, explicitly specified checks, rather than a reconstruction of the archived GMM-3 plots. It verifies the circular rates, the time-aligned radius and derivative, the exact eccentric-orbit solution and Fourier coefficients, the stationary-model bounds above, and the finite-window response in Eq.~\eqref{eq:finite_window_response}. For the stated HEO, composite-trapezoidal integration on 32768 uniform intervals differs from the exact Keplerian timing curve by less than 0.0089 ps at all sampled epochs, before any mid-range recentering.

To test the mechanism discussed for Fig.~\ref{fig:circular_budget_full}, an independent circular $i=92^\circ$ spacecraft starts at the ascending node on the $+X$ axis. In this coplanar external-body diagnostic, the Sun and both moons move prograde in the equatorial plane; their starting longitudes are $0$, $\pi/4$, and $2\pi/3$, respectively. It uses the manuscript's Mars/moon constants, a mean solar distance of 1.523679 au, and the explicit analytic-test value $GM_\odot=1.32712440018\times10^{20}\,\mathrm{m^3\,s^{-2}}$, not a new ephemeris fit. At 160 logarithmically spaced altitudes from 120 to 25000 km, doubling the intervals per revolution from 4096 to 8192 changes the Sun, Phobos, and Deimos half-amplitudes by at most $5.2\times10^{-7}$, $8.0\times10^{-8}$, and $4.7\times10^{-10}$ ps, respectively. Four further moon phases demonstrate phase-dependent non-monotonicity. These tests isolate geometry and finite-arc integration; they are neither the missing original phases nor a phase-complete worst-case envelope. The stable point-mass expression, full settings, CSV results, and plots are included in the accompanying verification note.

The archived figure files and manuscript report the GMM-3 calculations but do not include their original generating code, coefficient-file copy/checksum, full state arrays, or complete phase and averaging metadata. They therefore support the reported scale hierarchy, not an independently revalidated degree-selection certificate. In particular, the mismatch between the three-revolution LMO plot and the separate truncation benchmark is made explicit above instead of assigning an unsupported common interval. Recovering that archive, or replacing it with a fully documented rerun, is required for exact replication of every original curve. The new analytic and tide checks are independently reproducible without that archive.

\subsection{Optional ephemeris/software implementation}\label{subsec:software_example}

The transformations do not require a particular software package. One possible realization uses the JPL Development Ephemerides (for example DE440/DE441) together with a consistent satellite ephemeris \cite{ParkDE440}; another validated ephemeris representation can be substituted. SPICE is an optional interface for retrieving those inputs, not a reference-system or time-scale definition. Its ephemeris-time argument is conventionally expressed as TDB seconds past J2000 \cite{NAIFTime}. The input states and mass parameters must then be handled consistently with Eq.~\eqref{eq:tdb_compatible_scaling}; an onboard spacecraft-clock count is not proper time merely because a clock kernel is provided.

In such an implementation, spacecraft-and-planet kernels (SPK), planetary-constants kernels (PCK), frame kernels (FK), leap-seconds kernels (LSK), and spacecraft-clock kernels (SCLK) encode state, orientation, frame, civil-time conversion, and onboard-clock calibration information. The selected identifiers, versions, origin, time argument, and scaling must be recorded. None of these kernel types is needed for the prescribed circular and Keplerian diagnostics in this paper, and no IAU endorsement of a software implementation is implied.

\section{One-way and two-way radio/optical time transfer}\label{sec:time_transfer}

\subsection{BCRS light-time equation}\label{subsec:bcrs_lighttime}

All interplanetary light-time observables should be modeled in the BCRS.  For emission at \((t_1,\bvec{x}_1)\) and reception at \((t_2,\bvec{x}_2)\), the leading static first-post-Newtonian form is
\begin{equation}
 t_2-t_1=\frac{R_{12}}{c}+\sum_B\Delta^{\rm Sh}_{B,\,1PN}+\Delta_{\rm mov}+\Delta_{2PN}+\Delta_{\rm media}+\Delta_{\rm inst}+\Delta_{\rm ant},\label{eq:lighttime}
\end{equation}
where \(R_{12}=|\bvec{x}_2(t_2)-\bvec{x}_1(t_1)|\) and
\begin{equation}
\Delta^{\rm Sh}_{B,\,1PN}=\frac{2GM_B}{c^3}\ln\left(\frac{r_{1B}+r_{2B}+R_{12}}{r_{1B}+r_{2B}-R_{12}}\right).
\label{eq:auto:037}
\end{equation}
This is the usual Shapiro term in general relativity \cite{Shapiro1964,Moyer2003}.  The symbols \(\Delta_{\rm mov}\) and \(\Delta_{2PN}\) stand for moving-body/retarded-position and second-post-Newtonian light-propagation terms, respectively \cite{KopeikinSchaefer1999,Teyssandier2008}.  They are written explicitly here to emphasize that a local clock-retention demonstration is not, by itself, a complete sub-picosecond interplanetary observable model.

The Sun's Shapiro delay can reach tens to hundreds of microseconds near solar conjunction and dominates the leading relativistic light-time correction.  The Mars local Shapiro coefficient is
\begin{equation}
\frac{2GM_{\mars}}{c^3}=3.18\,\mathrm{ps};
\label{eq:auto:038}
\end{equation}
for grazing Mars-to-orbiter or Mars-to-Earth geometry the logarithmic factor can raise the Mars Shapiro term to tens of ps, which is above the formal $0.1\,\mathrm{ps}$ threshold.  Earth, Jupiter, and Saturn Shapiro terms are also retained for interplanetary links when the line of sight passes near the body.  Near solar conjunction, a grazing-Sun post-post-Newtonian scale such as \((15\pi/4)G^2M_\odot^2/(c^5R_\odot)\) is of order \(10^2\,\mathrm{ps}\).  Thus a one-way \(0.1\,\mathrm{ps}\) interplanetary link realization requires the standard moving-body and 2PN solar terms, together with media and hardware calibrations; these terms are not bounded by the Mars-local clock simulations in Sec.~\ref{sec:numerical_simulations}.

For a two-way coherent radio observable,
\begin{equation}
\rho=t_3-t_1=\Delta t_{1\to2}+\Delta t_{2\to3}+\Delta_{\rm transponder}+\Delta_{\rm ramp}+\Delta_{\rm media},
\label{eq:auto:039}
\end{equation}
with the event time $t_2$ solved iteratively.  The spacecraft proper time, oscillator phase, and transponder delay must be mapped to TCA and TDB consistently using Eq.~\eqref{eq:proper_tca} and Eq.~\eqref{eq:tca_tcb}.

\subsection{Mars Sagnac and rotating-station terms}\label{subsec:mars_sagnac}

For a Mars surface station and a Mars orbiter link, the first-order Mars Sagnac term is of order
\begin{equation}
\Delta_{\rm Sag}^{\mars}\sim\frac{\bvec{\Omega}_{\mars}\cdot(\bvec{r}_1\times\bvec{r}_2)}{c^2}.
\label{eq:auto:040}
\end{equation}
For a surface-to-areostationary link the maximum scale is
\begin{equation}
 |\Delta^{\rm surf-ASO}_{\rm Sag}|_{\max}\simeq\frac{\Omega_{\mars}R_{\rm eq}r_{\rm ASO}}{c^2}=5.5\times10^{-8}\,\mathrm{s}\simeq55\,\mathrm{ns},
\label{eq:sagnac_surface_aso_scale}
\end{equation}
Equations~\eqref{eq:auto:040} and
\eqref{eq:sagnac_surface_aso_scale} are rotating-frame
representations.  In a BCRS event-time solution, the same effect is
generated by evaluating the transmitting and receiving station states
at their distinct event times.  An explicit Sagnac correction must
therefore not be added a second time; it is used only in an equivalent
rotating-frame or reduced local formulation \cite{Moyer2003}.  The
surface-to-areostationary scale in
Eq.~\eqref{eq:sagnac_surface_aso_scale} corresponds to a
\(16\,\mathrm{m}\) one-way light-time effect, or about
\(5\times10^{5}\) times the \(0.1\,\mathrm{ps}\) timing gate.  The Sagnac term is therefore mandatory for precise Mars navigation.  For surface-to-LMO links, $\Omega_{\mars}R_{\rm eq}(R_{\rm eq}+h)/c^2$ gives a maximum scale near $10\,\mathrm{ns}$ for $h=300$--$400\,\mathrm{km}$, with actual values bounded by that scale and reduced by the visible station--spacecraft geometry.

Radio links must also model Earth troposphere, Earth ionosphere, solar plasma, spacecraft antenna phase centers, station coordinates, relativistic frequency transfer, ramped uplink frequencies, transponder ratios, and antenna mechanical delays.  Optical links reduce plasma effects but require atmospheric and pointing corrections.  These media and hardware terms will usually dominate the formal relativistic residuals unless carefully calibrated.

\section{Implementation workflow}\label{sec:implementation}

A flight-quality Mars timing implementation should be numerical-first.  The analytic formulae and diagnostic simulations in this paper identify the required terms and provide independent checks; they do not yet constitute a delivered operational Mars Time Ephemeris.  The reference product for operations should be a Mars Time Ephemeris and mission clock-transformation product generated from a single internally consistent set of data products, with a documented forward/inverse BCRS--MCRS residual.

\subsection{Step 1: choose conventions, constants, and data products}\label{subsec:workflow_step1}

The first step is to freeze the convention set.  The minimum data package is:
\begin{enumerate}
\item a selected planetary ephemeris, together with the satellite ephemeris needed to construct the Mars body-center worldline used by the gravity and rotation models \cite{ParkDE440};
\item documented time-conversion, body-state, orientation, frame-transformation, and onboard clock-calibration data in any suitable validated representation;
\item a Mars gravity field such as GMM-3 or MRO120D, including normalization convention, reference radius, $GM_{\mars}$, static coefficient covariance, and time-variable low-degree terms; for sub-picosecond surface-scale work, seasonal CO\(_2\)/loading coefficients must be modeled, monitored, or carried as explicit systematic errors \cite{Genova2016,Konopliv2016,PGDAGMM3};
\item a Mars body-fixed frame and rotation model based on IAU/WGCCRE conventions, with any mission-specific updates for precession, nutation, polar motion, and prime meridian \cite{Archinal2018};
\item topography/areoid and landed-site coordinates for surface clocks;
\item link models for antenna phase centers, hardware delays, troposphere, ionosphere, solar plasma, optical path delay, and transponder or coherent-link turnaround ratios.
\end{enumerate}
All constants must be stored in one configuration file, and the TDB-compatible or TT-compatible status of every $GM$, position vector, and time argument must be explicit.

\subsection{Step 2: generate the TCA--TCB/TDB transformation}\label{subsec:workflow_step2}

Compute the Mars barycentric energy function $\mathcal{E}_{\mars}(t)$ in Eq.~\eqref{eq:autoalign:001} on the ephemeris time grid, including the retained $c^{-2}$ and $c^{-4}$ terms of Eq.~\eqref{eq:tca_tcb}.  Numerically integrate
\begin{equation}
I_A(t)=\int_{t_0}^{t}\mathcal{E}_{\mars}(t')\,\dd t'
\label{eq:auto:041}
\end{equation}
with an adaptive high-order quadrature or a fixed-step Gauss/Jackson or Dormand-Prince dense-output integrator.  The quadrature tolerance should be chosen so that the accumulated numerical error is below $0.03\,\mathrm{ps}$ over the longest interpolation span.  Fit
\begin{equation}
I_A(t)=L_A(t-t_0)+P_A(t)-P_A(t_0)
\label{eq:auto:042}
\end{equation}
over the operational interval, and store both the mean rate $L_A$ and the periodic part $P_A(t)$ as Chebyshev coefficients or a dense tabulation.  The product should also report the maximum forward/inverse TCB$\leftrightarrow$TCA interpolation residual over the tabulated interval.  This paper gives the construction and Mars-local diagnostic residuals; the ephemeris-specific $L_A$, $P_A(t)$ table and its forward/inverse residual are future product outputs.  The position-dependent term $(\bvec{v}_{\mars}\cdot\bvec{r}_{\mars})/c^2$ must be evaluated at each event time, not absorbed into the mean rate.

\subsection{Step 3: define and realize the Mars surface scale}\label{subsec:workflow_step3}

Adopt $L_{\rm surf}^{\rm def}$ and paired epoch readings $T_{\mathrm{CA},0}$ and $T_{M0}$ for $T_M$, and separately document the selected areoid potential $W_0^{\mars}$ and $\delta L_0$ from Eq.~\eqref{eq:Lsurf_convention}.  For each landed clock compute
\begin{align}
\Delta W_i^{\rm def}(T)&\equiv W_i(T)-W_{\rm def}\nonumber\\
&=\big[W_i(T)-W_0^{\mars}\big]+c^2\delta L_0,\nonumber\\
\frac{\dd\tau_i}{\dd T_M}&=1-\frac{\Delta W_i^{\rm def}(T)}{c^2}+\cO(c^{-4}).
\label{eq:auto:043}
\end{align}
The first bracket contains topography, static harmonics, centrifugal potential, tides, seasonal gravity, loading, and local motion; the second term is the convention-to-areoid offset.  Uncertainty in both must be carried in the realized clock transformation.

\subsection{Step 4: propagate spacecraft states and integrate proper time}\label{subsec:workflow_step4}

For each orbiter, propagate the MCRS state with the same gravity and rotation model used for the clock transformation.  In the schematic force model below, SRP denotes solar radiation pressure:
\begin{equation}
\ddot{\bvec{X}}=\nabla U_{\mars}+\nabla U_{\rm tid}+\bvec{a}_{\rm SRP}+\bvec{a}_{\rm drag}+\bvec{a}_{\rm maneuvers}+\bvec{a}_{\rm rel}+\cdots .
\label{eq:auto:044}
\end{equation}
Then integrate the clock equation simultaneously,
\begin{equation}
\dot\tau=1-\frac{1}{c^2}\Big\{\frac{1}{2}V^2+U_{\mars}(\bvec{X})+U_{\rm tid}(\bvec{X})\Big\},
\label{eq:auto:045}
\end{equation}
The sign of the gravitational acceleration follows the positive convention $U=GM/r$: $\nabla U$ points inward. The gradient of the origin- and dipole-subtracted tidal potential already includes the differential third-body acceleration, so no duplicate $\bvec a_{\rm 3body}$ is added. Likewise, Eq.~\eqref{eq:exact_tide_model} already contains each moon's direct point-mass potential. Near a moon, a resolved gravity model replaces that source term (with the same origin/dipole subtraction); it is not added to the existing point-mass term.  The integration step should resolve the fastest retained clock line.  For 300-km LMO this means resolving a $\sim1.9\,\mathrm{h}$ orbital period and a $2n$ gravity-harmonic line; for highly elliptical orbits it means resolving periapsis passage.

\subsection{Step 5: solve light-time and measurement equations}\label{subsec:workflow_step5}

Form radio and optical observables in the BCRS.  Transform station and spacecraft states at their respective event times, solve emission, turnaround, and reception iteratively, and apply the retained Shapiro, moving-body, 2PN, media, antenna, and hardware terms.  Represent Mars rotation either through event-dependent BCRS station states or through an equivalent explicit Sagnac term, but not both.  The iteration tolerance should consume only a fraction of the $0.1\,\mathrm{ps}$ budget; a $3\times10^{-14}\,\mathrm{s}$ change in computed light time is consistent with the $0.03\,\mathrm{ps}$ numerical allocation used in Step~2.  The final observable should be verified against a tighter reference solution.

\subsection{Step 6: extract analytic retention tables and onboard approximations}\label{subsec:workflow_step6}

After the numerical reference model is built, fit $P_A(t)$ and $P_{\rm orb}(t)$ with sinusoids, Chebyshev polynomials, or short arc-wise series.  Apply the secular, maximum-rate, and finite-arc timing metrics in Eqs.~\eqref{eq:retention_rate_metric} and~\eqref{eq:retention_timing_metric}.  Terms below their individual gates must still be bounded in combination: use direct evaluation or a coherent worst-case sum, and use root-sum-square only when an explicit covariance model justifies the assumed correlations.

\section{Validation strategy}\label{sec:validation}

\subsection{Internal mathematical validation}\label{subsec:internal_validation}

The first validation layer is internal consistency.  For each simulated or propagated orbit, verify that the split
\begin{equation}
\dot q(t)=L_{\rm orb}+\dot P_{\rm orb}(t),\qquad \langle\dot P_{\rm orb}\rangle=0,
\label{eq:auto:046}
\end{equation}
closes over integer orbital periods.  For circular orbits, compare the numerically extracted $J_2$ and solar-tide amplitudes against the closed-form degree-2 and quadrupole-tide scalings plotted in Fig.~\ref{fig:circular_budget_full}.  For eccentric orbits, compare direct quadrature against Eq.~\eqref{eq:elliptic_exact} and its Bessel-function Fourier expansion, and verify convergence as the time step is reduced.  The BCRS--MCRS transformation should be checked by transforming Mars-center events forward and backward and confirming that the residual is below the interpolation tolerance.

A validation campaign should apply four independent checks.

First, the barycentric transformation should be validated at the Mars
center by comparing the numerical \(L_A\) and \(P_A(t)\) obtained from Eq.~\eqref{eq:mars_time_ephemeris_split} with an independent integration over the same planetary ephemeris.  The forward--inverse
\(\TCB\leftrightarrow\TCA\) residual should remain below the interpolation tolerance of the time ephemeris.

Second, the local proper-time integrator should be checked against
analytic limits: circular-orbit secular rates from Eq.~\eqref{eq:L_circ},
the polar \(J_2\) timing line of Eq.~\eqref{eq:J2_line}, and the
Keplerian eccentric-orbit solution of Eq.~\eqref{eq:elliptic_exact}
together with the Bessel-function Fourier expansion in Eq.~\eqref{eq:elliptic_series}.

Third, the gravity-field truncation should be validated by differencing
each truncated field against the adopted degree/order-120 GMM-3 reference reconstruction over the same prescribed trajectory and reporting both fractional-rate and integrated timing residuals.

Fourth, the complete observable model should be tested with real or
archival radiometric data by inserting the clock transformation into the
range/Doppler/light-time solution and verifying that estimated clock,
orbit, and media parameters do not absorb deterministic relativistic
signatures predicted by the model.

\subsection{Validation with existing Mars tracking data}\label{subsec:mars_tracking_validation}

Existing Mars missions can validate the dominant parts of the model even without optical clocks.  Long-arc Mars orbiter Doppler and range data are sensitive to the same gravity field, Mars orientation, and ephemeris quantities that enter the timing transformation.  Reprocessing tracking data with and without selected relativistic terms provides a differential test: the residual signatures should appear at the predicted periods and phases.  Useful cases include low-altitude mapping orbiters for $J_2$ and high-degree gravity signatures, areosynchronous or high-apogee relays for solar-tide signatures, and lander/orbiter range arcs for surface-site potential and Mars rotation consistency.  The validation observable is not the spacecraft clock alone; it is the full computed-minus-observed residual in the Deep Space Network (DSN) light-time, Doppler, and range measurement model \cite{Moyer2003}.

\subsection{Cross-comparison with Earth-based time scales}\label{subsec:earth_time_crosscomparison}

A Mars Time Ephemeris should be cross-compared with terrestrial products by evaluating $T_M-\TT$ and $\TCA-\TDB$ over multi-year intervals.  The proper time of a clock on the selected areoid should then be formed with Eq.~\eqref{eq:areoid_tt_bridge} before comparison with Ref.~\cite{AshbyPatla2025}; the coordinate quantity $T_M-\TT$ must not be identified directly with a physical areoid-clock difference.  High-precision comparison also requires the same averaging interval, constants, ephemeris time argument, reference potential, and Earth time ephemeris, together with phase and amplitude tests of the Mars annual, Earth--Mars synodic, and position-dependent BCRS--MCRS terms.

\subsection{Future direct clock validation}\label{subsec:future_clock_validation}

The most decisive validation would use a stable Mars lander clock and a second clock in a well-tracked orbiter.  The lander proper time must first be corrected to $T_M$ with Eq.~\eqref{eq:auto:014}; otherwise site height, local potential, seasonal gravity, and $\delta L_0$ enter the measured rate difference.  A 300-km orbiter differs from $T_M$ by $4.56\,\mu\mathrm{s/day}$ and carries an $\simeq87\,\mathrm{ps}$ Mars-$J_2$ line.  An areostationary clock differs by $9.13\,\mu\mathrm{s/day}$ and has solar-tide signatures of $0.28\,\mathrm{ps}$ at mean distance and $0.37\,\mathrm{ps}$ at perihelion.  A highly elliptical relay produces a maximum local Keplerian excursion near $0.35\,\mu\mathrm{s}$.  Realization requires calibrated delays, media corrections, and accurate spacecraft dynamics.

\section{Discussion}\label{sec:discussion}

The Mars time-scale problem is not merely a relabeling of the lunar or terrestrial problem.  Mars is dynamically dominated by heliocentric motion, has a substantially eccentric orbit, has large topographic relief, and has a gravity field whose seasonal low-degree variations are part of the clock environment.  Consequently, the common Mars--Earth rate term is large and strongly time dependent; the areoid realization is geophysically nontrivial; and practical relay architectures span low circular orbits, high circular relays, moon-related trajectories, and highly elliptical orbits.  A single closed-form expression is therefore less useful than a controlled workflow: define the frames, integrate the coordinate-time transformation numerically, define the surface scale conventionally, integrate proper time along each trajectory, and extract analytic approximations only after the numerical reference model is available.

A second important point is the distinction between formal and realized accuracy.  The formal post-Newtonian terms can be carried below $0.1\,\mathrm{ps}$, but realized Mars navigation will often be limited by ephemeris errors, media delays, spacecraft clock noise, solar plasma, trajectory maneuvers, antenna phase centers, gravity-field errors, and surface-site coordinates.  The value of the formal model is that it prevents avoidable relativistic biases from entering the measurement system and provides a common language for cross-mission interoperability.

For standardization, we recommend that a Mars time convention contain five elements: (i) a formal MCRS/TCA definition as a Mars-centered IAU-style coordinate system; (ii) a conventional value and epoch for the Mars surface scale $T_M$; (iii) a reference gravity/areoid/rotation model for realization, not for the definition itself; (iv) a Mars Time Ephemeris giving $\TCA-\TCB$, $T_M-\TDB$, and associated periodic terms; and (v) a validation protocol tied to DSN observables, Mars orbiter tracking, and future clock payloads.  Mars solar/civil time scales should remain separate from TCA and $T_M$ because they serve scheduling and daylight conventions, not relativistic equations of motion.

The residuals reported in this paper are not a complete
mission covariance.  A realized Mars time system must also propagate uncertainties in the planetary ephemeris, Mars orientation, gravity-field coefficients and covariance, seasonal low-degree gravity, surface coordinates and areoid height, spacecraft non-gravitational accelerations, maneuvers, antenna phase centers, radio/optical media delays, transponder/hardware delays, and clock noise.  The deterministic threshold \((5\times10^{-18},0.1\,{\rm ps})\) should therefore be interpreted as a model-retention criterion, while the realized clock and navigation performance must be established by a joint orbit, clock, and media-delay estimation covariance.

\section{Conclusions and recommendations}\label{sec:conclusions}

We have constructed a step-by-step relativistic time-scale framework for Mars surface operations and Mars-orbiting clocks.  The central chain is
\begin{equation}
\BCRS/\TCB \longrightarrow \MCRS/\TCA \longrightarrow T_M \longrightarrow \tau_{\rm surface},\;\tau_{\rm orbiter},
\label{eq:auto:047}
\end{equation}
with TDB and TT connected through the standard IAU definitions.  The MCRS/TCA construction follows directly from the IAU BCRS/GCRS post-Newtonian transformation specialized to the Mars ephemeris origin, in the same formal spirit as the LCRS/TCL construction now used for lunar time-scale work \cite{Turyshev2025Time,Turyshev2026Cislunar}.  The Mars surface scale $T_M$ is best treated as a conventional coordinate-time scaling, analogous in spirit to TT.  A physical areoid clock is related to it through the explicit convention offset $\delta L_0$ and the site correction in Eq.~\eqref{eq:auto:014}; it is not rate-matched to $T_M$ merely by lying on a geophysically defined areoid.

The quantitative hierarchy, summarized in Table~\ref{tab:term_hierarchy}, is clear.  Mars areoid clocks differ from Earth geoid clocks by hundreds of microseconds per day, with a large annual modulation \cite{AshbyPatla2025}.  In the leading monopole/Keplerian rate estimates relative to the Mars surface scale, a 300-km circular orbiter loses $4.56\,\mu\mathrm{s/day}$, a 1000-km relay loses $1.90\,\mu\mathrm{s/day}$, a representative $300\times17000\,\mathrm{km}$ highly elliptical relay gains $7.02\,\mu\mathrm{s/day}$, an areostationary clock gains $9.13\,\mu\mathrm{s/day}$, and a Deimos-distance relay gains $9.52\,\mu\mathrm{s/day}$.  The latitude-sampling Mars-$J_2$ timing line is $\simeq87\,\mathrm{ps}$ at 300-km LMO, $\simeq21.5\,\mathrm{ps}$ at Phobos distance, $\simeq6.7\,\mathrm{ps}$ at areostationary radius for a polar-sampling trajectory (not an exactly stationary equatorial clock), and $\simeq5.4\,\mathrm{ps}$ at Deimos distance.  The solar quadrupole tide is negligible for LMO but must be retained in high relay orbits; the mean-distance values $0.28$--$0.45\,\mathrm{ps}$ become $0.37$--$0.60\,\mathrm{ps}$ at Mars perihelion and therefore set the retain/discard decision.  Highly elliptical relay orbits require direct quadrature because their local kinematic-plus-monopole timing residuals are at the $0.1$--$1\,\mu\mathrm{s}$ level.

The recommended implementation is numerical-first.  Build a Mars Time Ephemeris from the adopted planetary and satellite ephemerides; define $T_M$ by the conventional constant $L_{\rm surf}^{\rm def}$; propagate spacecraft with a mission-approved Mars gravity, rotation, and force model; integrate proper time along the spacecraft trajectory; solve one- and two-way light-time observables in the BCRS, representing Mars station motion consistently, with an equivalent Sagnac scale of $55\,\mathrm{ns}$ for surface--ASO links; and then extract analytic retention tables for onboard or simplified applications.  The diagnostics demonstrate the required hierarchy, while the independent checks document analytic and tidal tests with explicit settings. The remaining original-figure provenance limits in Sec.~\ref{subsec:independent_checks} must not be mistaken for a completed numerical certification.  The formal relativistic model should retain every term above $5\times10^{-18}$ in fractional frequency or $0.1\,\mathrm{ps}$ in one-way timing amplitude, while the combined omitted contribution should be evaluated directly or bounded with a justified correlation model.

The results should be interpreted as a relativistic reference-system and model-retention framework, not as a final operational Mars time ephemeris. The latter requires a single adopted planetary ephemeris, a Mars orientation model, a gravity-field and seasonal-loading model, spacecraft orbit-determination trajectories, calibrated link delays, and a clock/orbit/media covariance. Within that scope, the present analysis identifies the deterministic Mars-local clock terms that exceed the \(5\times10^{-18}\) fractional-rate or \(0.1\,{\rm ps}\) one-way timing gates for the stated diagnostic models and provides a reproducible procedure for turning Mars-centered coordinate time into proper time for surface and orbital clocks.

The leading Mars-local limitation not closed by the static deterministic simulations is time-variable low-degree gravity.  Seasonal low-degree coefficients can map to surface fractional-rate perturbations of order $10^{-18}$--$10^{-17}$, depending on coefficient amplitude, latitude, normalization, and the adopted seasonal model \cite{Genova2016}, and can therefore exceed the $5\times10^{-18}$ gate.  A sub-picosecond Mars surface-scale realization must model, monitor, or empirically bound seasonal CO\(_2\) loading and related atmospheric/solid-body terms rather than treating the static GMM-3 residuals as the complete surface error budget.

The circular zero-drift orbit, $r_{\rm TA}=5082.6$ km for the adopted $T_M$ convention, provides a useful design reference but removes only the secular offset; harmonics, tides, and physical site realization remain necessary. This framework provides the technical basis for interoperable Mars timing: landed geodetic clocks, one-way and two-way radio/optical links, relay navigation, inter-spacecraft synchronization, future Mars GNSS-like services, and interplanetary time transfer connected consistently to TT/TDB.  The next step toward a standard is a full numerical Mars Time Ephemeris generated from a selected planetary ephemeris and Mars gravity/orientation solution, together with open validation cases and residual tests against existing Mars tracking data.

\section*{Acknowledgments} 
We thank Todd A. Ely and Ryan S. Park of JPL for their comments, suggestions, and constructive criticism on this manuscript. The work described here was carried out at the Jet Propulsion Laboratory, California Institute of Technology, Pasadena, California, under a contract with the National Aeronautics and Space Administration.
\textcopyright 2026. California Institute of Technology. Government sponsorship acknowledged.

\appendix

\section{Useful order-of-magnitude formulae}\label{app:scaling_formulae}

The following formulae are useful for rapid budget checks:
\begin{align}
1\,\mathrm{ps} &\leftrightarrow 0.299792458\,\mathrm{mm},\qquad 1\,\mathrm{ns}\leftrightarrow0.299792458\,\mathrm{m},\\
5\times10^{-18}\times86400\,\mathrm{s} &=0.432\,\mathrm{ps},\\
\frac{g_{\mars}h}{c^2} &=4.13\times10^{-17}\Big(\frac{h}{1\,\mathrm{m}}\Big),\\
L_{\circ} &=\frac{3}{2}\frac{GM_{\mars}}{rc^2},\\
|P_{J_2}|_{\rm polar} &\simeq \frac{3}{8}\frac{GM_{\mars}R_0^2J_{2\mars}}{c^2r^3n},\\
|P_{\odot,2}| &\simeq \frac{3}{8}\frac{GM_\odot r^2}{c^2R_{\mars\odot}^3\omega_{\rm samp}}.
\label{eq:autoalign:002}
\end{align}
Here $\omega_{\rm samp}$ is the frequency at which the trajectory samples the solar quadrupole: it is of order the mean motion for a simple inertial circular-orbit estimate and equals $\Omega_{\rm syn}$ for a body-fixed areostationary clock.  Use the instantaneous Mars--Sun distance in a flight product and direct quadrature for high-eccentricity orbits.

\section{Recommended model-retention policy by regime}
\label{app:retention_policy}

Table~\ref{tab:retention_policy} summarizes evidence-based guidance for the diagnostic cases in Secs.~\ref{sec:orbit_regimes} and~\ref{sec:numerical_simulations}.  Degree/order 120 denotes the static GMM-3 reference reconstruction; a lower truncation is valid only after both maximum-rate and finite-arc timing residuals have been checked for the specified trajectory.

\begin{center}
\refstepcounter{table}
\label{tab:retention_policy}
{TABLE~\thetable.} Model-retention guidance by Mars operational regime.
\par\vspace{0.5ex}
\renewcommand{\arraystretch}{1.05}
\begin{tabular}{lll}
\hline\hline
\parbox[t]{0.16\textwidth}{\raggedright Regime} & \parbox[t]{0.37\textwidth}{\raggedright Mars field} & \parbox[t]{0.32\textwidth}{\raggedright External/moon terms}\\
\hline
\parbox[t]{0.16\textwidth}{\raggedright Surface / lander} &
\parbox[t]{0.37\textwidth}{\raggedright Static field, topography, stated areoid, and the $W_0^{\mars}-W_{\rm def}$ convention offset; time-variable low-degree gravity is a realization term.} &
\parbox[t]{0.32\textwidth}{\raggedright Solar tide, seasonal CO$_2$, loading, site height, and actual site velocity.}\\[0.8ex]

\parbox[t]{0.16\textwidth}{\raggedright 300--400 km LMO} &
\parbox[t]{0.37\textwidth}{\raggedright Use degree/order 120 as the reference.  For the displayed 300 km case, the RMS screen remains above the nominal rate gate to about degree 40, while the three-revolution plot exceeds 0.1 ps at degree 10. The separate degree-16 benchmark does not certify that arc; verify both gates on the selected trajectory.} &
\parbox[t]{0.32\textwidth}{\raggedright Solar tide is below the timing gate for the displayed short arc; third bodies remain in the dynamical and ephemeris models.}\\[0.8ex]

\parbox[t]{0.16\textwidth}{\raggedright 1000 km relay} &
\parbox[t]{0.37\textwidth}{\raggedright Degrees in the tens remain relevant in the RMS screen, but no finite-arc minimum is established here; verify directly against degree/order 120.} &
\parbox[t]{0.32\textwidth}{\raggedright Solar tide is small in the displayed circular-orbit budget; retain third bodies for dynamics.}\\[0.8ex]

\parbox[t]{0.16\textwidth}{\raggedright Phobos-/Deimos-distance relay} &
\parbox[t]{0.37\textwidth}{\raggedright Mars $J_2$ remains above the timing gate; no universal high-degree minimum is established.} &
\parbox[t]{0.32\textwidth}{\raggedright Retain the solar quadrupole at Deimos distance and moon self-potentials for close operations.}\\[0.8ex]

\parbox[t]{0.16\textwidth}{\raggedright Areostationary} &
\parbox[t]{0.37\textwidth}{\raggedright Retain the constant $J_2$ and longitude-dependent $C_{22}/S_{22}$ rate offsets; verify any truncation against the reference for the actual libration, inclination, and eccentricity.} &
\parbox[t]{0.32\textwidth}{\raggedright Retain the solar quadrupole with instantaneous Mars--Sun distance and synodic sampling.}\\[0.8ex]

\parbox[t]{0.16\textwidth}{\raggedright HEO relay} &
\parbox[t]{0.37\textwidth}{\raggedright Use direct quadrature along the osculating trajectory.  The archived benchmark reports a degree-8 timing residual of $0.0375\,\mathrm{ps}$; it does not certify the maximum-rate, orbit-mediated error, or another arc.} &
\parbox[t]{0.32\textwidth}{\raggedright Retain the solar tide near apoapsis and use moon-centered models for close approaches.}\\
\hline\hline
\end{tabular}
\end{center}


%

\end{document}